# Prediction of cancer driver genes and mutations: the potential of integrative computational frameworks

*Authors*

Mona Nourbakhsh[1,†], Kristine Degn[1,†], Astrid Saksager[1], Matteo Tiberti[2], Elena Papaleo[1,2]*

†These authors have contributed equally to this work

*Affiliation*

[1] Cancer Systems Biology, Section for Bioinformatics, Department of Health and Technology, Technical University of Denmark, 2800, Lyngby, Denmark

[2] Cancer Structural Biology, Danish Cancer Society Research Center, 2100, Copenhagen, Denmark

*Abstract*

The vast amount of sequencing data presently available allow the scientific community to explore a range of genetic variables that may drive and progress cancer. A myriad of predictive tools has been proposed, allowing researchers and clinicians to compare and prioritize driver genes and mutations and their relative pathogenicity. However, there is little consensus on the computational approach or a golden standard for comparison. Hence, benchmarking the different tools depends highly on the input data, indicating that overfitting is still a massive problem. One of the solutions is to limit the scope and usage of specific tool. However, such limitations forces researchers to walk on a tightrope between creating and using high-quality tools for a specific purpose and describing the complex alterations driving cancer. While the knowledge of cancer development increases every day, many bioinformatic pipelines rely on single nucleotide variants or alterations in a vacuum without accounting for cellular compartment, mutational burden, or disease progression. Even within bioinformatics and computational cancer biology, the research fields work in silos, risking overlooking potential synergies or breakthroughs. Here, we provide an overview of databases and datasets for building or testing predictive tools for discovery of cancer drivers. We introduce predictive tools for driver genes, driver mutations, and the impact of these based on structural analysis. Additionally, we suggest and recommend directions in the field to avoid silo-research, moving in the direction of integrative frameworks.

*Keywords*

Driver Genes, Driver mutations, Protein Structural Analysis, Pathogenicity, Predictive Tools, Computational Research, Cancer

*Introduction*

Cancer is a group of diseases characterized by uncontrolled cell growth and tumor formation, a result of genomic alterations. The hallmarks of cancer provide a scaffolding framework for interpreting the processes involved in cancer development [1–3]. Genomic alterations in cancer can include mutations, chromosomal and epigenetic changes. These genomic alterations occur in the so-called driver genes which promote cancer development and progression (**Figure 1A**), conferring selective growth advantages to the cancer cells [2–5].

The mutations that confer a selective growth advantage to the cancer cells are driver mutations whereas mutations with no effect on the selective growth advantage of the cell are called passenger mutations [5] as illustrated in **Figure 1B**. Driver mutations can impact protein structural stability and/or function, leading to gain or loss of function (**Figure 1C**) [6–8].

In silico predictions of driver genes and mutations rely on high-quality data, which can be obtained through next-generation sequencing technologies [9]. Data quality and handling should be closely examined when used for tool development [10]. To date, many computational tools and frameworks have been proposed for the prediction of driver genes, driver mutations, and to assess the structural impact of protein variants. These methodologies are important for assessing the pathogenic potential of mutated variants and designing the most suitable downstream experiments for validation [11]. However, without robust tools and frameworks to analyze the exponentially growing pool of data, we cannot infer meaningful biological interpretations, representing another bottleneck. Thus, solid bioinformatics pipelines are pivotal for increasing our knowledge of tumorigenesis and ultimately, drug target discovery.

In this review, we provide an overview of relevant datasets and databases for cancer research and a subset of tools for prediction of driver genes or driver mutations, as well as structure-based frameworks. The aim is to address the advantages and limitations of the tools to give a comprehensive understanding of the current status. We also discuss possible future directions within each of these arms of research and collectively. We excluded tools developed before 2015 to limit our initial search and to decrease the risk of including outdated tools due to lack of maintenance. This review is targeted towards both developers and users who may benefit from understanding the biological impact of the technical features of the tools.

*Datasets and Databases to study cancer genes and mutations*

Ten years ago, the main holdback for developing high-quality tools to differentiate drivers and passengers was the lack of high-quality curated datasets [12]. Since then, several datasets and databases have been developed based on manual or automated curation. Although manual curation poses the advantages of incorporating expert-based knowledge and critical judgment on a subject, it may suffer from omission of important discoveries missed or dismissed by the curators. A database built upon literature mining may challenge the speed and ease of future updates of the database, especially if the data mining is based on dictionaries that are not standardized [13]. However, even with a perfectly unbiased curated dataset, the way the data is sampled and balanced is also a potential source of error in driver classification. This is especially accentuated in the study of cancer variants since the number of passenger is much higher than the one of driver mutations [14,15]. To solve this imbalance, two general strategies are used: i) remove passenger mutations[16] or ii) increase the driver mutations, either using subsampling or synthetic additions in the training set [17]. However, similar approaches do not remove the issues related to introduce bias and the down-sampling strategy might miss important information [18]. An alternative could be to build or benchmark tools based on datasets specifically curated for the purpose [19]. As an example, a benchmark dataset containing driver genes with both passenger and driver variants mapped is now available [19]. To overcome biases, datasets should be carefully evaluated for their origin, heterogeneity, data balance, data processing, and curation method. Some of the biases that should be evaluated is the representation of cancer types, tumor stages, the clinical profile of the patients, and demographic composition [20].

Pan-cancer initiatives, (**Table S1**), such as The Cancer Genome Atlas (TCGA) [21] and the International Cancer Genome Consortium (ICGC) [22] have accelerated cancer research. In parallel, increased understanding of cancer drivers has led to the creation of datasets annotated with information on specific driver genes (**Table S2**) or mutations (**Table S3**). For example, OncoKB [23] is a precision oncology database that includes more than 3000 genomic alterations in around 400 cancer-associated genes and incorporates therapeutic implications, offering guidance for clinicians and cancer researchers [23].

To study the structure of a protein, the traditional resource is the Protein Data Bank (PDB) [24] containing experimentally solved structures. With the rise of reliable de-novo model databases, researchers have gained access to predicted structures of high quality. Notably, the AlphaFold2 database [25,26] contains predictions of protein structures for an ever-growing fraction of the human proteome. Other databases are also rapidly becoming household names such as the ESM metagenomic atlas [27]. Overall, these databases should be seen as the source of starting structures to study mutation impact and not the end point of a structural study.

*Driver Gene Predictions*

Since the detection of the first cancer genes, the field of cancer genomics has exploded. This has led to the discovery of more than several hundred driver genes and continues to be a major goal of cancer research [28]. For example, Bailey et al. [29] performed a pan-cancer analysis of 33 cancer types to discover cancer driver genes and mutations using 26 different tools. They found 299 driver genes of which 59 had not previously been reported by six other pan-cancer studies or in the Cancer Gene Census (CGC) [30]. Moreover, they predicted 3442 driver mutations using both sequence- and structure-based approaches [29].

In terms of biological function, driver genes encode proteins that participate in various cellular processes such as cell proliferation, cell survival, and genome maintenance [5]. Predicting the role of driver genes in relation to the cancer hallmarks could contribute to active reversal of the disrupted pathways [1]. Most human cancers develop because of only a small subset of alterations occurring in the driver genes [5]. Fortunately, in the last decade, computational methods have accelerated driver gene discovery. These driver gene prediction tools rely on different principles, methods, and data input (**Figure 2**). Here, we divide these tools into different categories based on the main principle that the method is built upon, i.e., network construction, machine learning, multi-omics data integration, and mutational information. We also review those tools that specifically focus on tumor suppressor, oncogene, and dual role gene prediction. Additionally, we discuss the challenges and limitations of the tools and the field.

*Interaction Network Construction*

Network-based approaches use network data and aim at modeling the role and impact of each gene in the network [31]. In these networks, nodes represent genes, and edges represent interactions between genes [32,33]. Additionally, these methods employ the concept of influence where the genes with the greatest influence in the network are likely the ones driving carcinogenesis [34–36].

Examples of tools utilizing the influence concept are iMaxDriver [36], GenHITS [34], KatzDriver [35], and DriverGroup [37]. These tools use interaction and/or gene expression data, on which a network is constructed, allowing for driver gene discovery through the influence concept. What distinguishes KatzDriver is the network construction. While the nodes and edges still represent genes and interactions, respectively, the nodes are categorized as either a transcription factor or non-transcription factor (mRNA). This division effectively constructs a transcriptional regulatory network. Thus, the regulatory interactions cover both transcription factor-transcription factor and transcription factor-mRNA interactions, calculating the relative gene level impact in the regulation as a measurement of the driver gene status [35]. Two features of DriverGroup [37] worth highlighting is the detection of groups of driver genes rather than individual drivers and its ability to detect both coding and non-coding driver gene groups. Despite the essential role of non-coding genes in cancer [38], few tools focus on detecting non-coding driver genes. These studies are challenged by the increased search space of the non-coding genome compared to the coding genome and the low number of known driver genes with non-coding alterations [20], exemplifying data imbalance and ascertainment bias.

Other network-based tools integrate the network and gene expression data together with additional sources of -omics data such as mutation, copy number variation, and DNA methylation data. For instance, AMARETTO [39], DriverFinder [40], CBNA [32], PRODIGY [41], and LNDriver [42] fall into this category. AMARETTO includes genes that have copy number or DNA methylation alterations as potential cancer driver genes. These potential driver genes are connected to their regulated targets. The regulatory modules created on individual cancer types are finally connected to a pancancer network in which driver genes are found [39]. DriverFinder [40] accounts the influence of gene length on the predictions, a distinctive characteristic compared to most other tools. DriverGroup [37] and CBNA [32] predict both coding and non-coding drivers. Finally, a notable feature of PRODIGY [41] is its ability to predict driver genes at a patient-specific level.

Finally, other network-based tools employ the concept of random walks to investigate the relationship between genes in a network. These tools also overall use mutation, gene expression, and network data as inputs but differ in their subsequent application of random walks on the constructed networks. Examples of them are Driver_IRW [33], Subdyquency [43], MEXCOwalk [44], and RLAG [45]. MEXCOwalk [44] exploits driver modules, which contains cancer-causing genes acting together in functional pathways, a concept similar to DriverGroup [37] . Moreover, a notable feature of Subdyquency [43] and RLAG [45] lies in their integration of information on subcellular localization. They rely on the principle that proteins can only interact if they are located in the same compartments [43,45].

Network-based approaches pose both advantages and disadvantages. They evaluate the genes on a network level which provides a holistic view. As cancer comprises many genes interacting in a network and not genes acting in isolation, these tools offer an effective way to represent this mechanism. On the other hand, these methods mainly predict coding drivers, overlooking the non-coding drivers. Additionally, they mainly consider interactions between transcription factors and target genes and exclude other types of interactions relevant as cancer driving mechanisms [32–36,40] .

*Machine Learning*

A recent review by Andrades and Recamonde-Mendoza points out an increasing interest in machine learning methods for driver gene prediction based on different features and algorithms [20].

A relatively underexplored field within driver gene prediction is at the intersection between network-based and machine learning approaches, i.e., graph-based machine learning [20]. MoProEmbeddings, for example, involves a novel node embedding procedure where genes in a network are represented by combining moment and propagation embeddings. The node embeddings are input to four different binary classifiers, i.e., logistic regression, random forests, support vector machines (SVMs) and gradient boosting. Each of them can be used to predict the class label of genes as cancer driver genes or not in a supervised way [46].

Another supervised approach is DriverML [47] which combines machine learning and a weighted score test. The weights represent the functional impacts that different mutation types (missense, nonsense, splice site, frameshift indel, and in-frame indel) have on protein function, quantified through a machine learning algorithm [47].

A handful of tools employ the functional impact of mutations on the protein product. They aim to better predict lowly recurrent mutated genes and genes that are mutated in the later stages of tumorigenesis [48,49]. sysSVM2 [50] is another supervised approach based on SVMs [20]. In sysSVM2, molecular and systems-level features of canonical drivers define a training set. The output of four different one-class SVMs are combined into a single score for each gene as a reflection of its resemblance to the canonical drivers. One advantage of one-class SVMs is that they overcome the class imbalance issue [20]. sysSVM2 allows for prediction of driver genes at the single patient level [50], a characteristic that is also available with driverR [51]. driveR uses a multi-task learning model to obtain cancer type-specific probabilities of genes being drivers or non-drivers. The features used in the model are derived from somatic genomics data, ANNOVAR [52] annotations, Phenolyzer [53] gene scores, and membership of genes to cancer-related KEGG pathways [54]. Multi-task learning was also implemented in MTGCN [55]. This seems a promising strategy for cancer type-specific modeling to detect cancer-specific driver genes which otherwise may not be predicted using pan-cancer models [20].

Finally, neural networks are also represented among the machine learning tools for prediction of cancer driver genes, although to a low extent. For example, FI-Net [56] uses an artificial neural network to estimate functional impact scores of genes by using features from genomics, transcriptomics, and epigenomics data sources. Estimating background FIS distribution and finding driver genes in clusters separated by multi-omics features highlight some of the novelties of this method [56]. DeepDriver [57] predicts driver genes by a convolutional neural network (CNN). Here, the convolution is performed through the combination of mutation-based features and gene similarity networks where the functional impact of mutations and gene expression similarities are learned simultaneously [57]. Finally, DeepCues [58] is a deep learning model that uses convolutional neural networks for cancer type classification and prediction of driver genes. DeepCues integrates somatic and germline variants, and insertions/deletions [58].

The performance of the above mentioned machine learning tools are dependent on i) the quality of the training data, ii) the amount of available data for training or validation, iii) curation of positive sets of known driver genes which are difficult to define as a gene may be a known driver in one cancer (sub)type, but there may not be absolute certainty that it is in another, and iv) creation of negative sets of non-driver genes which are likewise difficult to define. Additionally, obtaining cancer type-specific known driver genes is challenging, creating known positive sets too small to yield reliable results. Some cancer types do not have enough data available to generate a fair training set. These issues can be overcome through applying pan-cancer models instead [20,59]. However, the importance of cancer type-specific algorithms is worth highlighting. For instance, in a comparison among predicted driver genes in different cancer types, markedly different candidate driver genes have been found among non-organ-related cancer types [60].

*Multi-omics Data Integration*

Other tools integrate additional layers of -omics data besides gene expression, mutation, and network data, namely copy number variation and methylation data. These two additional layers of complexity allow for a deeper understanding of the underlying cancer mechanisms and genomic alterations occurring due to chromosomal rearrangements and epigenetic changes. For example, OncoIMPACT [61] and iPDG [62] integrate gene expression, interaction network, and mutation data and copy number variation data. Overall, these tools perform driver gene analysis similarly to the above-mentioned network-based approaches that integrate multi-omics data. OncoIMPACT evaluates the impact of genomic mutations by using interaction networks to associate the mutations with transcriptomic changes and modules of patient-specific deregulated genes. Moreover, OncoIMPACT allows for patient-specific driver gene prediction. In iPDG, copy number variation and mutation data are integrated in a single dataset which is then mapped to the interaction network. Correspondingly, DriverSubNet maps differentially expressed and mutated genes to a protein interaction network. The subsequent steps deviate from other implementations since DriverSubNet creates subnetworks for each mutated gene whose dysregulation is evaluated through enrichment analysis [63].

One attribute excluded from the above-mentioned tools are epigenetic alterations which can be analyzed through a DNA methylation layer. Despite the known fact that DNA methylation aberrations contribute towards cancer progression, a limited number of tools distinguish DNA methylation driver changes from passenger changes, MethSig focuses on promoter hypermethylation as an inactivating mechanism of tumor suppressors and predicts DNA methylation driver genes through a novel statistical framework [64]. iEDGE integrates gene expression profiles with (epi-)DNA alterations such as somatic copy number alterations, DNA methylation, mutation, or microRNA regulatory networks. iEDGE performs differential expression analysis to find cis and trans genes associated with the (epi-)DNA alteration, carries out pathway enrichment analysis of the significantly differentially expressed cis and trans gene sets, and predicts cis driver genes of the alteration using a mediation analysis [65]. frDriver also integrates functional impact scores with gene expression and mutation data. frDriver is built on Bayesian probability and multiple linear regression models where the goal is to identify protein regions that regulate gene expression levels and have high functional impact potential [66].

An advantage of methods combining various -omics data lies exactly in the data integration. Multiple data types provide many sources of information that can provide a more comprehensive picture of the underlying mechanisms, potentially improving the performances. On the other hand, the integration of different data types is a challenge in itself. Different datasets may be obtained from different sources, leading to lack of standardization and unintended confounders. Moreover, it is not guaranteed that the required data types from each integrative tool are available for the user, potentially limiting their applicability. Additionally, methylation and copy number variations are largely understudied compared to mutational impacts. Hence, integrating these additional layers can contribute towards novel insights within cancer biology.

*Mutational Information*

A vast amount of cancer driver gene prediction tools utilizes and analyze mutation data to identify driver genes. Many tools compare observed mutation frequencies with a background mutation model to discover driver genes, such as OncodriveCLUSTL [67] and ActiveDriverWGS [68]. OncodriveCLUSTL is a sequence-based unsupervised clustering algorithm that identifies clusters of somatic mutations in selected genomic elements. ActiveDriverWGS predicts candidate non-coding drivers among genomic regions of cis-regulatory modules through evaluation of single nucleotide variants and indels in each region compared to a region-specific expected mutation model. An accurate background mutation models is difficult to create due to tumor heterogeneity and these methods are also challenged by driver genes with a low mutation frequency.

QuaDMutNetEx combines mutual exclusivity and network approaches. In QuaDMutNetEx, somatic mutations are used to find mutual exclusivity patterns and biological networks are used as context for the observed data. QuaDMutNetEx is suited for discovery of driver genes with low mutation frequency [69]. Another example is MaxMIF which integrates somatic mutation and protein-protein interaction data using a maximal mutational impact function. MaxMIF is based on connections in the protein-protein interaction network and provides an estimate of the mutational impact function for each pair of genes [70].

Finally, a class of approaches is based on Bayesian frameworks. For example, cDriver integrates signatures of selection of somatic mutations (single nucleotide variants and short indels) in a novel framework at three levels: the population level (population recurrence), the cellular level (cancer cell fraction), and the molecular level (functional impact) [71]. MADGiC incorporates various mutation features for driver gene prediction at once. These features are mutation frequency, mutation type, gene-specific features known to affect background mutation rates, functional impact scores, and spatial patterning of mutations [72]

The advantage of mutation-based tools is their applicability to additional mutation datasets [32]. However, driver genes are prone to diverse types of genomic alterations, which risk to be overlooked. Moreover, many mutation-based tools predict driver genes by comparing observed mutation frequencies with a background mutation model. Methods that apply scores of functional impacts might be able to mitigate these problems since they are not only based on the recurrency of the mutations [48]

*Prediction of Tumor Suppressors, Oncogenes, and Dual Role Genes*

Cancer driver genes are classified into three categories: tumor suppressors (TSGs) which normally limit cell growth, oncogenes (OCGs) which normally promote cell growth, and dual role genes which exhibit both tumor suppressive and oncogenic behavior depending on the cellular context [5,73,74]. OCGs are activated by gain-of-function mechanisms and often encode transcription factors, chromatin remodelers, signal transducers, apoptosis regulators, growth factors and their receptors [75]. In contrast, TSGs are activated by loss-of-function and broadly encode intracellular proteins that regulate specific stages of cell cycle or cell proliferation, checkpoint-control proteins that regulate cell cycle arrest, apoptosis-inducers, and genes involved in DNA repairing [76]. Besides TSGs and OCGs, dual role genes have also been reported, contributing to the cancer complexity. Shen et al. have systematically characterized, in a pan-cancer manner, dual role genes through database search and text mining [77]. They found that majority of dual role genes were either transcription factors or kinases. Additionally, they performed network analysis of the dual role genes and discovered that they are often hub genes in protein-protein interaction networks. Similarly, Datta et al. reviewed four tumor suppressors pRb, PTEN, FOXO and PML, which can also display oncogenic functions [73]. Distinguishing between these categories of driver genes is important for gaining a deeper knowledge of the biological context of cancer. Below we have described a subset of driver gene prediction tools that offer a classification of driver genes in TSGs or OCGs.

GUST [78] and 20/20+ [79] use a random forest model to predict TSGs, OCGs tumor suppressors, oncogenes, and passenger genes based on 10 and 24 features, respectively. The model included in GUST is based on differences in selective patterns of the three gene classes [78]. Similarly, 20/20+ integrates features of positive selection to predict the different gene classes from small somatic variants [79].

DORGE [80] and LOTUS [59] are based on machine learning algorithms. DORGE uses two elastic net-based logistic regression models (DORGE-OCG and DORGE-TSG) which include 75 features broadly divided into mutational, genomic, epigenetic, and phenotypic features. Not many tools consider epigenetic data, a property that sets apart DORGE from similar methods. Furthermore, the two separate classifiers allow for the prediction of dual role genes. LOTUS uses one-class SVM and integrates mutation frequency, functional impact, and pathway-based information in terms of protein-protein interaction networks in its framework and allows for the prediction of driver genes in both a pan-cancer and a cancer type-specific setting using a multitask learning strategy.

In 2020, we also contributed to this field with Moonlight [81] which predicts OGCs, TSGs and dual role genes using multi-omics data. The idea behind MoonlightR is that we have a primary layer where we integrate gene expression data and information about biological processes to detect genes that we defined as oncogenic mediators. However, since gene expression data alone predicts many genes which are not all driving the cancer phenotype, at least one second layer (or more) of evidence is needed. Such secondary layers can for example be DNA methylation, copy number variation, mutation, or chromatin accessibility data. If evidence for deregulation of an oncogenic mediator is provided through this secondary layer, the user can retain those oncogenic mediators as our final set of driver genes. This allows a mechanistic explanation of the activation or inactivation of the oncogenic mediators. Moonlight allows for the prediction of oncogenic mediators using either an expert-based approach or a machine learning approach. The expert-based approach utilizes patterns of opposing cancer-related biological processes for the predictions. For example, if these two processes are apoptosis and cell proliferation, then those differentially expressed genes with a positive effect on apoptosis and a negative effect on cell proliferation are deemed putative tumor suppressors and vice versa for putative oncogenes. The machine learning approach predicts the oncogenic mediators using a random forest model. We recently automatized the integration of a secondary mutational layer in a new function available in the second version of Moonlight, Moonlight2 [82]. The oncogenic mediators containing at least one driver mutation are retained as the driver genes. Besides classification of mutations, the new implementation allow to predict effects of mutations on the transcriptional, translational, and protein structure/function level, thereby aiding mechanistic explanations of the deregulated driver genes, ultimately illustrating an integrative computational framework (**Figure 3**).

As evident, many driver gene prediction tools exist, however, only a limited number of these tools classify driver genes into TSGs, OCGs and dual role genes. These three gene classes drive cancer development through different biological mechanisms, and more tools distinguishing these categories, in a context-dependent manner, are needed to increase our understanding of cancer biology.

*Summary and Recommendations*

As evident from the above sections, numerous driver gene prediction tools have been developed. However, a consensus is missing due to predictions of driver genes by different tools lacking consistency. This leads research groups to use their own developed methods [83]. Different research efforts showed a small consensus among predicted driver genes from different methods [56,60,79]. While the true number of driver genes in a cancer (sub)type is unknown, discovering an adequate number of driver genes is vital. Tools suffering from under-selection predict too few driver genes, potentially overlooking important genes. On the other hand, tools suffering from over-selection risk a large false positive rate and complicate subsequent experimental studies.

Validating driver gene prediction tools is challenging as a gold standard of known driver genes and a universally accepted standard for this procedure does not exist, which is a limitation for benchmarking studies [61,62,70,71,79]. Most studies evaluate the performance of driver predictors by comparing the overlap between the predicted driver genes and driver genes listed in the COSMIC Cancer Gene Census (CGC) and the Network of Cancer Genes (NCG) databases. CGC is an expert-curated database of 719 genes that drive human cancer [30]. NCG is a manually curated database containing 2372 genes with cancer driving functions [84]. Moreover, the original list of known driver genes by Vogelstein et al. is often used [5]. However, CGC and the list by Vogelstein et al. are embedded in NCG. This can pose a challenge when the datasets of known driver genes are used in both development and validation steps of a tool, e.g., using the CGC to train your model and subsequently validating it on the NCG will lead to overfitting of your model. One should be careful in the study design to ensure these lists are not intertwined and thereby prevent overfitting during development and validation of driver predictors.

Tokheim et al. established an evaluation framework to assess and compare method performances, circumventing the use of a gold standard. This framework includes five components. The authors established that a method that can discover many of the driver genes from CGC and those predicted by other methods fulfill two criteria of good performance [79]. These guidelines were, for example, used by Parvandeh et al. to compare their tool EPIMUTESTR with state-of-the-art methods [85].

Each method has its advantages and disadvantages and consequently, combining the results from multiple methods would aid the discovery and evaluation of critical driver genes [60,79]. It appears that while most studies generally utilize one tool, newer studies are beginning to incorporate two or three approaches [84].

The majority of the driver gene prediction tools are cohort-level methods, meaning they predict driver genes across patient cohorts. However, these methods often fail to identify rare driver genes present in only few patient and mapping the predicted driver genes back to the patients leaves many patients with few or no driver genes. Additionally, these predictions are challenging to use in the clinic as they predict driver genes for the whole cohort instead of for individual patients [50]. For these reasons, patient-level tools have recently emerged which are valuable for clinical strategies [41,50,51].

**Driver Mutations Prediction**

Oncogenesis originates in a few key driver mutations [86,87]. Identification of specific driver mutations could inform further studies of functionality to find novel druggable targets [88,89]. Today, a plethora of tools aims to predict driver mutations. Yet, a consensus on the approach for analysis is still absent. Even tools developed to discover pathogenic mutations across diseases are routinely used to identify cancer driver mutations. Collectively, the tools are termed variant effect predictors (VEPs). The initially published tools within this field relied on frequency measurements by identifying mutations that appear significantly more often than in the background model [72]. While these frequency measurements indicate an evolutionary association, the approach lacks sensitivity toward healthy human variation [90]and misclassifies 26-38% of known pathogenic mutations [91]. To overcome this, most tools in use today rely on a combination of genomic features, such as proximity to splice sites or transcription factor binding sites, evolutionary features such as conversion-based frequency measurements, coding data such as information regarding germline variations or somatic mutations, physicochemical properties and protein domains [92,93].

To understand and discuss these tools, we here present a subset of the current tools as categorized based on the computational approach (**Figure 4**). The first group is based on frequency and statistics, the second group on supervised machine learning algorithms, and the third group on unsupervised machine learning algorithms.

*Statistical Methods*

The statistical scoring methods are primarily based on frequency measurements and are often considered functional impact scores based on conservation. The aim is to measure the frequency of a mutation compared to a background model.

To identify driver mutations based on an evolutionary conservation score, examples of available tools are Protein Variation Effect Analyzer (PROVEAN) [94], which predicts the functional effect of alterations using pairwise alignment. Sorting Tolerant From Intolerant (SIFT) [95,96] predicts if any amino acid substitution affects protein functionality. This functionality is classified based on an evolutionary conservation score using multiple sequence alignment that calculates the probability of the substitution. Such statistical tools are limited by the annotation methods and conservation metrics that vary between them and seldom account for allele specificity or functional information [97]. The performance of the tools is evaluated by comparison to known pathogenic mutations, which may introduce ascertainment bias due to a limited amount of available annotations. Each tool utilizes its own scoring measure, which is not compatible with other scores, or any physical measurement [97]. Furthermore, as with driver genes, the accuracy of the background model for driver mutations can be a limitation due to tumor heterogeneity. An additional constraint is the recurrence of mutations, which are often not included in background mutation rates [98]. The authors provided an alternative based on probabilistic models estimating mutability per nucleotide.

DriverML relies on frequency measurements and further applies the scores to create mutational clusters that account for missense, nonsense, splice site, frameshift, and in-frame indel mutations [47]. Other useful approaches to the study of pathogenic mutations relies on evolutionary information, such as GEMME [99] and DeMaSk [100]. GEMME is based on evolutionary-informed conservation where the quantification of the impact of variants takes into account global similarities and is not limited to a single amino acid change. DeMaSk predicts the variant impact using the fitness impact which is estimated based on a linear model of data from deep mutational scans and an asymmetrical amino acid substitution matrix. Here, a loss of fitness indicates a negative impact on the protein caused by the variant. The asymmetrical substitution matrix takes the direction of the mutation into account, thereby differentiating between substitution to and from residues.

*Supervised Machine Learning Methods*

The advantages of using supervised learning to classify a particular mutation as either driver or passenger is the opportunity to include a large number of parameters. However, these tools are limited by the data on which they are built - a good performing supervised learning model requires a balanced dataset and considerations of the class imbalance problem.

*Random Forest:* A well-known tool residing in this category is REVEL [101]. REVEL is a method for predicting the pathogenicity of missense variants. REVEL is built on a random forest model with 1000 binary classification trees where the input is a range of several. REVEL outputs a score between 0 and 1 reflecting the average classification from each tree. REVEL has good performances and is routinely used as the benchmarking standard for new driver mutation prediction tools.
Another tool utilizing the random forest algorithm is M-CAP [91] which classifies clinical pathogenic mutations, focusing on variants of unknown significance (VUS). The input features consist of preexisting pathogenicity scores, publicly available biological features, and a metric for evolutionary constraint. The preexisting scores mirror the scores used in REVEL, while also including unsupervised methodologies such as Eigen [102]. The publicly available features cover base-pair, region, gene, and amino acid conservation obtained from - among others – PhyloP [103] and the standard PAM250 and BLOSUM62 matrices. The final feature is metrics for evolutionary constraint on amino acid residues in coding genes. In a comparison among different tools, REVEL featured consistent better performances on ClinVar, TP53 and PPARG data, illustrating how there may be an optimum in the number of features or ensemble methodologies [104].
CHASMplus aims at scoring the oncogenic impact of a driver missense mutation specifically by cancer type. The tool is constructed from TCGA data from 32 cancer types as a classifier that differentiates between the driver and passenger missense mutations. The tool is built on many features of which the most impactful feature is HotMaps 1D [105]. This indicates that structural information could be utilized to find and assess driver mutations. Another example of utilizing structural information is DENOGEN2 [106], aiming to predict deleterious single nucleotide variants.
In summary, a range of tree-based methods exist, proving that a combination of evolutionarily based tools is beneficial and that these can be combined with other types of descriptors.

*Support Vector Machines (SVMs) & Logistic Regression*: Another approach to differentiate between driver and passenger mutations is SVMs. SVMs bring an advantage when applying to high-dimensional data, and the model is less prone to find local minima compared to other machine learning methodologies. However, SVMs lack transparency in the weight of the model, there is no probability estimate, and the computational time is very high which is not optimal considering the rapidly increasing amount of available data. A tool within this category is CADD [15,97,107]. Essentially, CADD aim is to score any possible human SNV or indel event and to integrate several different approaches to the annotation of genomics to build a robust classifier. CADD was initially built as a support vector model but has migrated to use a logistic regression classifier over time. This was due to an increase in dataset size resulting in increased training time.

*Hidden Markov Models (HMMs)*: One example is the FATHMM framework to predict genome-wide pathogenic point mutations. The framework consists of FATHMM-cancer [108], FATHMM-MKL [16] and FATHMM-XF [109] where the underlying HMM is built on protein domain annotations. Each version of the tool adds features, i.e., functional annotations in the -MKL version and probability of predicted pathogenicity in -XF.

*Gradient Boosting*: The advantage of boosting algorithms is the transparency of the calculations and its resilience to overfitting. However, some disadvantages include heightened sensitivity to outliers and historical issues with upscaling. AI-Driver predict the driver status of somatic missense mutations, using gradient-boosted trees to combine scores from other tools, which include machine learning algorithms mentioned above [89]. BoostDM applies a similar underlying algorithm, considering gene-tumor combinations, and in particular, cancer mutations available for that specific combinations, which constitute a driver mutation set. A second passenger mutation set is generated by means of a stochastic process. Mutations are annotated with a number of mutational features. These datasets are used to train a gradient boosting classifier. This process is repeated for several gene-tumor combinations, and the resulting collection of models constitutes BoostDM [17]. A different approach using the same algorithmic approach to predict driver frameshift indels is PredCID. Here, the developers generated eight different biological features categorized into gene, DNA, transcript, and protein levels and implemented a XGBoost (eXtreme Gradient Boosting) classifier to distinguish the driver and passenger frameshift indels [110]. Another approach is CScape-somatic, which uses a forward selection through a greedy search algorithm, relying on thirty features differentiated over five feature groups: conservation, local mutation frequency, distance from gene features and two related to sequence: GC content and sequence uniqueness [92]. All these tools have been published since 2020, illustrating an abundance of methods available that may be combined to gain more insights.

*Neural Network - Regression*: The main advantages of neural networks are their computational affordability and ability to find connections in a flexible manner. However, they are a black box that may challenge the interpretation of the output, especially when the labeled training data may not be fully annotated from a biological standpoint. The MutPred framework consists of two tools: MutPred-LOF [111] and MutPredInDel [112]. They were developed by applying an identical methodology, but on different datasets. MutPred-LOF predicts pathogenic and tolerated loss-of-function variants (frameshift and stop codons) while MutPredInDel covers pathogenic and tolerated non-frameshifting insertion/deletion variants. They were developed by the ensemble of 100 bagged two-layer feed-forward neural networks, including evolutionary, structural, and functional features for loss-of-function genetic variants. The evolutionary feature is a Position-Specific Scoring Matrix (PSSM) built on the training data. The structural information is calculated using a vector quantization framework where the structural information is dimensionally reduced and clustered and accounts for secondary structures as well as signal peptide and transmembrane sections. The functional features included information on motifs, enzyme activation, and post-translational modifications. For every mutation input into the MutPred framework, the model returns a score between zero and one - variants with higher scores are more likely to be pathogenic. In summary, all the supervised machine learning methods combine existing tools and occasionally physicochemical and structural information to gain a higher true positive rate in the respective classifiers.

*Unsupervised Machine Learning Methods*

A comparatively small range of unsupervised tools compared to supervised tools exist. This lack of abundance is connected to the assumption that differentiating driver and passenger mutations is a binary classification problem. When introducing unsupervised machine learning, the answer is unlikely to be binary. The limitations of these tools are the risk of clustering mutations without guarantee that the clusters represent driver status. A classic unsupervised machine learning approach is used to build PrimateAI [113] which aims to identify pathogenic missense mutations. PrimateAI approaches the driver mutation question by creating an artificial neural network, but shies away from using any precomputed functionality such as the supervised learning classifiers did. Rather, it inputs sequences and bases all calculations on sequence homology to other species. One non-neural network example is Eigen [102]. Here, an eigenvector weighted score is calculated for the identification and annotation of disease variants. The aim is to combine multiple binary classifiers of unknown reliability by calculating the covariance matrix, then blocking the binary classifiers, which is i) an evolutionary conservation block, ii) a regulatory information block, and iii) an allele frequency block. These are used to construct a rank-one matrix of which the eigendecomposition is taken. The Eigen score becomes the score calculated as the eigenvector weighted sum of the annotations. A different approach has been taken by Allodriver [114] relying entirely on structural data. Allodriver aims to identify and prioritize driver mutations based on known allosteric and orthosteric sites derived from three-dimensional structures. The model is constructed as a combination of random forest and feed-forward neural network models on an oversampled dataset of driver mutations. Another example of a combined tool is GenoCanyon [115] which relies on the posterior probability of conserved regions and biochemical annotations to annotate each position in the human genome. Lastly, EVE [116] relies on protein sequence and its evolution to estimate protein variant pathogenicity. The authors start from the observation that supervised models may introduce inflated accuracy as they are measured against a clinical label of pathogenicity, which may not capture the true value. EVE is an unsupervised generative model generating a score for each mutation in a protein on a scale from benign to pathogenic (0-1). The model relies on encoding multiple sequence analysis, allowing the computation of an evolutionary index which is used as input for the gaussian mixture model separating benign and pathogenic variants. Unsupervised methods, in general, rely on more basic biological annotations such as structure, which could potentially remove some biases that may have been included in previous tools for prediction of driver mutations.

*Summary and Recommendations*

Even though all the tools mentioned above are variations of the same objective, a consensus on a benchmarking protocol has not been reached. One benchmarking study was conducted by Livesey et al. who compared the performance of 46 predictors towards a dataset of deep mutational scanning data from 31 experiments [117]. The studied mutations were not associated with any one disease. They found the predictive performance to vary considerably between tools. They suggest that these differences stem from known limitations in predictive models: i) re-use of training data and ii) ascertainment bias, performing well on heavily studied genes. Another benchmarking study was done by Chen et al. [118] who aimed to compare the performance of different algorithms on five datasets to benchmark the tools. They found that tools specifically designed to deal with cancer performed better than disease agnostic tools. However, it is also evident that performance of the tool changed significantly depending on the selected dataset. This illustrates the fundamental limitations of these tools, which are connected to either: i) a limited amount of known and annotated driver mutations, which restricts training set size, or ii) the scope of the tool. Rather than designing a comprehensive tool with the purpose of predicting any driver mutation, it may be preferable to design a set of tools that may predict specific effects of these mutations such as regulatory impact or protein loss- or gain-of-function. Rogers et al. suggests that the fundamental idea of dividing mutations into drivers and passengers is a constraint as it would be more prudent to ask why a mutation is a driver rather than if [119]. Another constraint is the annotation of identified driver mutations. A mutation could be annotated as a driver, however, this could be the result of insufficient or incomplete data if said mutation is only a driver in one context but a passenger in another. Tools trained on this annotation may introduce underlying biases. A solution could be achieved by stratifying the prediction not only down to a cancer type, but at a cancer subtype level and including information regarding the genes in which the mutation dwell and the associated cellular pathways and regulatory networks [118].

Future directions within this field should focus on cancer subtype-specific resources, including information regarding the placement of the mutation in a gene context, or applying new types of co-evolutionary analysis, and understand the molecular mechanism related to the mutations. A possibility to solve some of these issues is to move towards structure-based frameworks to understand cancer mutations.

*Structure-based Frameworks to Study the Effects of Cancer Mutations*

One way to study alterations driving cancer in the coding regions is to model these in the protein structure. For mutations with unknown consequence, this can be useful to predict their effect and understand what changes they impart to the protein structure; for mutations with known effects, structural studies help derive a mechanistic explanation of their consequence. This is achieved by identifying patterns in the structural changes in terms of stability disruption and changes in conformation at functional sites [120]. The primary amino acid sequence does not contain information regarding the three-dimensional folding of the protein. Accordingly, functional sites are often difficult to identify from the sequence alone, and mutations that are far from each other in the protein sequence may be adjacent in the structure. After folding, seemingly distant substitutions can impact functionality by means of orthosteric and even allosteric effects [114]. The study of protein structure has historically been challenging, and experimental structure determination is still expensive and time-consuming. To overcome this challenge, a range of structure prediction methods have been proposed over the years, with varying degrees of success [121]. Most prominently, AlphaFold2 [25] is currently the overall most accurate de novo prediction method [121]. We here present a subset of the current tools as categorized based on the computational approach (**Figure 5**): i) networks based on graph theory both within a structure and in interfaces between structures, ii) mutational clusters identifying structural-functional sites with high mutational burden, iii) scores based on distance measures, and finally, iv) scores based on Gibbs unfolding/folding free energy changes upon mutation.

*Networks*: When protein structure is encoded as a network, nodes represent protein residues and edges, or connections, represent some measure of interaction between pairs of them. The advantage of describing a protein structure as a network is a great simplification of the protein structure and the possibility to use graph-theoretical methods. Tools using this approach include HotMaps [105], HotCommics [122], e-Driver3D [123], and PyInteraph2 [124]. HotMaps aims to identify missense mutational hotspots considering the three-dimensional structure [105]. Other methods identify positions in the protein structure particularly enriched with cancer mutations and identifies spatially close groups of such residues as components in a residue interaction graph. An example in this category is FASMIC which was developed on a combination of experimental structures and homology models [125]. Within the graph, these methods use the Girvan-Newman algorithm to find communities, which is a group of residues connected almost exclusively to each other, and they can subsequently calculate if a community is more frequently mutated than expected by chance. HotCommics is developed for somatic cancer missense mutations and was developed on experimentally solved structures [122]. The e-Driver3D method, on the contrary, is cancer-specific and is based on protein-protein interaction networks and aims to analyze the mutation distribution of specific interaction interfaces [123]. The mutations were extracted from TCGA data and annotated from ENSEMBL, and the structures were reported in the PDB. eDriver3D calculates a ratio between the number of residues involved in the protein-protein interface and the total number of residues in the protein by applying networks that score the probability of interaction. The protein-protein interfaces are defined with a distance of 5 Å between residues on different chains. The potential of this approach was consolidated by Cheng et al. who applied protein-protein networks to cancer mutations to shed light on the high mutational burden in protein-protein interfaces by use of networks, which they experimentally validated [120]. PyInteraph2 applies graph theory to ensemble of structures with a broader scope [124]. PyInteraph has been used to study mutational effects, design variants for proteins, study biomolecular complexes, or understand the effect of post-translational modifications. Here, the creation of the network relies on intra- and intermolecular interactions between residues or side-chain contacts. The nodes are the residues, and the edges are the non-bonded interactions weighted based on the occurrence of the interaction in the ensemble. The tool allows the user to set thresholds for the network and supply a graph analysis module to, among others, identify nodes, hubs, connected components and other metrics.

The future steps in structural graph networks may include graph attention neural networks [126]. Approaches like these where the graph nodes capture predictive features while the edges are weighted by co-evolution align with the driver mutation predictive development and deployment of sophisticated machine learning approaches.

*Mutational Clusters:* A number of structural tools aims to identify and assess mutation clusters on the structure. Clusters are often defined based on a specific measure of spatial distance between residues in the structure, which is designed to be representative of the overall distance between residues, yet there is no consensus on how such measure is defined. For example, Mutation 3D aims to identify clusters of mutations to find driver genes. It is a cancer-specific tool that applies to somatic missense mutations [127]. Mutation 3D is developed on a combination of structures from PDB and homology models and uses hierarchical clustering with complete linkage, considering distances between alpha-carbons, to identify mutational clusters. Two tools aiming to identify mutational clusters and interpret their functional role are Hotspot3D [128] and CLUMPS [129] which both rely on somatic cancer mutations and PDB structures with high sequence identity to the chosen target. In Hotspot3D, the clusters are found based on the minimum distance between atoms in pairs of residues. The functional impact is subsequently annotated using ClinVar data [130]. Hotspot3D predictions have been experimentally validated, and the tool has been utilized within other applications, e.g., FASMIC [125]. In CLUMPS, clusters are found based on proximity as the clusters are calculated based on centroids as pairwise euclidean distance, where 5 Å is deemed the max distance of the contact. The identification of cancer mutations relies on testing if a cluster is more mutated than expected by chance. Furthermore, it is a possibility that one residue is present in multiple clusters. The method was also experimentally validated [131]. Finally, ASTID evaluates the three-dimensional spatial patterns of human germline and somatic variation, which is not necessarily cancer-specific. ASTID relies on both neutral germline variants, disease-causing germline variants, and recurrent somatic variants. This approach quantifies spatial distributions of protein-coding mutations to create clusters. Ripley's K is used to quantify the spatial heterogeneity between variants, the distance is measured using Euclidean distance, and distances over 45 Å were excluded based on their functional domain disparity [132].

The application of cluster-based methodologies identifying hotspots can be a valuable tool to differentiate drivers and passengers. However, the methodologies have two main limitations: i) they rely on a background model, whose shortcomings have previously been described, and ii) they apply different distance measurements and thresholds, making comparison difficult. However, mutational clusters by application of clustering algorithms hold great promise due to their efficiency in identifying alterations at functional sites because of mutational agglomeration.

*Distance-Based Scores*: Evaluating genes and mutations in the three-dimensional space can also be expressed as a score. One example of such an approach is StructMAn [133], which scores the change of conformation in terms of distance to a ligand. The aim is to classify nonsynonymous single nucleotide variants as either Quasi-WT (no apparent change of interactions), quasi-null (complete loss of interactions), or edgetic (specific loss of some interactions) based on set score thresholds. The score is based on a model from PDB using BLAST and the score is calculated as a weighted score that takes the quality of the model, the sequence identity, and the distance from the residue to the ligand molecule into account. An entirely different approach to a score is used in 3DTS (three-dimensional tolerance score) which is a score for missense mutations to describe functional constraints [134]. 3DTS works by taking variants with mutations within 5 Å from a feature defined in Uniprot and assessing the probability that the three-dimensional site is intolerant to a missense mutation. This is calculated using a model, which accounts for the differences among loci in the rates of neutral missense variation due to the genetic code, differential sample availability, and regional mutation rates. The main limitation of the distance-based scores is the reliance on single amino acid substitutions to drive a change in conformation resulting in a functional impact. These scores, however, may create a more comprehensive picture if incorporated with the mutational clusters.

*Prediction of changes in stability upon mutation*: An alternative to the tools described above is structural evaluations of alterations using free energy calculations [7], a task that has been recently streamlined thanks to high throughput workflow for mutational scans in silico, such as MutateX [135] and RosettaDDGPrediction [8]. The overall idea is to estimate the change of energy upon one or more mutations to assess the functional impact in terms of stability and interaction. They are commonly applied to understand the impact of structural alterations. The impact is expressed as changes in Gibbs free energy, and interpretation of these values depends on, for example, the expected accuracy of the prediction and whether the tool assesses stability changes or changes in binding energy. The advantage of these tools is how the changes in Gibbs free energy are physical measurements that can be compared to real world experimental data rather than an arbitrary score. These methods still suffer from limitations, partially because of the limited conformational sampling they are able to carry out, because of their intrinsic bias due to their unbalanced training set and of their sensitivity respect to the used structure. These free energy calculations were build based on experimentally found structures but does also apply to homology models [136] and de-novo models such as Alphafold2 [137]. Another possible scenario is to use deep learning models to predict free energy changes. An example is the recently developed RaSP [138] is a protein stability prediction tool capable of conducting a saturation mutagenesis within a few minutes. RaSP is created as a deep learning counterpart to Rosetta predictions with similar performance.

*Summary and Recommendations*

Much like the identification and assessment of driver genes and driver mutations, the area of structural assessments is fast developing. Particularly with the advent of Alphafold2, the largest limitation to these structural assessments, the limited available structures, may be a constraint of the past. This may pave the way for a range of new tools employing different machine learning algorithms and even more sophisticated clustering. To ease the barrier of entry to structural studies, further development may include options to visualize and analyze missense variants in a protein sequence and structural space for a set of variants found in the general population including protein-protein interactions, PTMs, and functional features as seen in MISCAST [139]. With accessibility may arise new ideas and applications from other computational science fields. To further strengthen the structural framework aiming to funnel -omics data into a tangible protein outcome, novel tools could take a leaf out of the driver mutation ensemble methodology. The aim of such an ensemble could be to describe the structural alterations resulting from the mutational clusters rather than individual assessments. This could lead to understanding the collective impact on function and binding, both using energy changes and distance scoring as well as considering the relative impacts of the mutations in the clusters such as compensation or synergistic effects. Creating an annotation and classification system for variants may lower the barrier of entry in interpreting protein structure studies as well as help prioritize experimental validation [11]. Yet, one limitation that should be addressed in future tools is the current modeling of proteins in a static conformation. Both the experimentally solved structures and the predicted structures rely on a single conformation of a protein. A protein's functionality is most likely dependent on its interaction with ligands and macromolecules, rendering them dynamic entities. One way to mitigate this limitation is to use molecular dynamic simulations to generate a representative ensemble of protein conformations in solution [11]. During the last couple of years, our group has been developing the first steps towards a more generalized structure-based framework to assess cancer variants on proteins involved in cancer hallmarks [11,140–143]. These studies created the foundation for a Multi-layered assessment of Variants by Structure for proteins framework, MAVISp [144]. MAVISp is an integrative module-based framework building a reproducible protocol to systematically study structural alterations.  MAVISp creates an end-to-end framework that can be applied to a single three-dimensional structure and its complexes or an ensemble of structures. The framework initially identifies known cancer mutations via COSMIC [145] and cBioPortal [146], but can also be supplied on a specific set of mutations from a particular research project [147]. MAVISp then identifies possible structures and known interactors of the protein within the structure selection and interactome modules. The following modules are stability, estimating the mutational impact on protein stability compared to the wildtype measured in changes in Gibbs free energy, local interactions, estimating the mutational impact on the interaction, and long-range effect, by estimating the allosteric free energy resulting from the perturbation of any residue and finally post-translational modifications. The point is to gain a thorough and rounded understanding of a protein. Additionally, MAVISp can handle data from structural ensembles and use them for the mentioned modules and add functional dynamics to the toolkit overcoming the limitations of using only a representative structure for a protein of interest (**Figure 6**).

*Discussion*

The field of predictive tools in cancer genomics has made significant progresses, but there are still several challenges that need to be overcome.
One major question is whether we can trust the predictions made by the aforementioned tools. While it is possible to predict possibly pathogenic mutations, the mechanism of action behind genetic alterations is not fully understood [148,149] . Another limitation is the trade-off between annotating every possible mutation and scoping the tools so narrow that only one context is investigated.
It should be noted that computational approaches serve as a vital starting point to elucidate the underlying mechanisms of cancer biology. These methods greatly reduce the number of genes and mutations to be tested experimentally and thereby decrease the experimental load, cost, and time associated with such tasks. Despite the importance of prior computational studies, wet lab validations should always follow to confirm the findings.
Not all tools are continuously maintained which can prevent steady use. This illustrates another challenge - the need to update the tool based on biological knowledge and computational resources available. Since many tools integrate existing biological knowledge from various databases, as seen for example with training sets in machine learning methods, these tools may benefit from routinely revisions and updates. While such knowledge is continuously updated, failing to incorporate novel findings and data dynamically in the tools may potentially decrease their performance and longevity. Moreover, the tools require maintenance in terms of software. Most of the tools are available as command line programs and Python and R packages. These platforms are also regularly updated, and consequently, the user may experience problems when installing and applying the tools due to incompatibility between the user's programming environment and the software requirements of the tool under which it was built. Besides maintenance, solid documentation is an important factor for successful usage. Thus, to circumvent obsoletion of a tool, developers should consider dynamic incorporation of biological knowledge, periodically test their tool when its underlying software is updated and ensure sound documentation including a suggested virtual environment or work towards community-driven efforts in tool development and maintenance.
Distinguishing between cancer driver and passenger genes and mutations is another challenge [150]. Fortunately, this obstacle can largely be addressed through the driver gene, driver mutation and structure-based frameworks described in this review, especially considering the fast-moving advancement of the field. Collectively, these predictive tools allow for analyses of the structural impact of the predicted driver alterations. Hence, we propose studying these fields collectively rather than individually, with each field serving as the input to the next, thereby promoting integration and avoiding research silos. Such a workflow would include initial driver gene prediction, subsequent driver mutation prediction in the driver genes, and finally, a structural assessment of the impact of mutations (**Figure 7**). This could in the future enable a transition from -omics analyses to drug discovery, repurposing, and development.

*Acknowledgements*

EP group has been supported by Hartmanns Fond (R241-A33877), Leo Foundation (LF17006), Carlsberg Foundation Distinguished Fellowship (CF18-0314), NovoNordisk Fonden Bioscience and Basic Biomedicine (NNF20OC0065262). EP group is also part of the Center of Excellence in Autophagy, Recycling and Disease (CARD), funded by the Danish National Research Foundation (DNRF-125).

*Contributions )*

Conceived and designed the review contents: K.D., M.N., E.P., Visualization: K.D., M.N., Wrote the review: M.N., K.D., M.T., A.S., E.P.

*References*

1. Hanahan D. Hallmarks of Cancer: New Dimensions. Cancer Discov 2022; 12:31–46
2. Hanahan D, Weinberg RA. Hallmarks of cancer: the next generation. Cell 2011; 144:646–674
3. Hanahan D, Weinberg RA. The hallmarks of cancer. Cell 2000; 100:57–70
4. Hahn WC, Weinberg RA. Modelling the molecular circuitry of cancer. Nat Rev Cancer 2002; 2:331–341
5. Vogelstein B, Papadopoulos N, Velculescu VE, et al. Cancer Genome Landscapes. Science 2013; 339:1546–1558
6. Gerasimavicius L, Liu X, Marsh JA. Identification of pathogenic missense mutations using protein stability predictors. Sci Rep 2020; 10:
7. Stein A, Fowler DM, Hartmann-Petersen R, et al. Biophysical and Mechanistic Models for Disease-Causing Protein Variants. Trends Biochem Sci 2019; 44:575–588
8. Sora V, Otamendi Laspiur A, Degn K, et al. RosettaDDGPrediction for high-throughput mutational scans: from stability to binding. Protein Science 2023; 32:e4527
9. Morash M, Mitchell H, Beltran H, et al. The role of next-generation sequencing in precision medicine: A review of outcomes in oncology. J Pers Med 2018; 8:
10. Hardwick SA, Deveson IW, Mercer TR. Reference standards for next-generation sequencing. Nat Rev Genet 2017; 18:473–484
11. Fas BA, Maiani E, Sora V, et al. The conformational and mutational landscape of the ubiquitin-like marker for autophagosome formation in cancer. Autophagy 2021; 17:2818–2841

12. Gonzalez-Perez A, Deu-Pons J, Lopez-Bigas N. Improving the prediction of the functional impact of cancer mutations by baseline tolerance transformation. Genome Med 2012; 4:
13. Learned K, Durbin A, Currie R, et al. Barriers to accessing public cancer genomic data. Sci Data 2019; 6:
14. Tokheim C, Karchin R. CHASMplus Reveals the Scope of Somatic Missense Mutations Driving Human Cancers. Cell Syst 2019; 9:9-23.e8
15. Rentzsch P, Schubach M, Shendure J, et al. CADD-Splice—improving genome-wide variant effect prediction using deep learning-derived splice scores. Genome Med 2021; 13:
16. Shihab HA, Rogers MF, Gough J, et al. An integrative approach to predicting the functional effects of non-coding and coding sequence variation. Bioinformatics 2015; 31:1536–1543
17. Muiños F, Martínez-Jiménez F, Pich O, et al. In silico saturation mutagenesis of cancer genes. Nature 2021; 596:428–432
18. Lawrence MS, Stojanov P, Mermel CH, et al. Discovery and saturation analysis of cancer genes across 21 tumour types. Nature 2014; 505:495–501
19. Raimondi D, Passemiers A, Fariselli P, et al. Current cancer driver variant predictors learn to recognize driver genes instead of functional variants. BMC Biol 2021; 19:1–12
20. Andrades R, Recamonde-Mendoza M. Machine learning methods for prediction of cancer driver genes: a survey paper. Brief Bioinform 2022; 23:1–19
21. Hutter C, Zenklusen JC. The Cancer Genome Atlas: Creating Lasting Value beyond Its Data. Cell 2018; 173:283–285
22. Zhang J, Bajari R, Andric D, et al. The International Cancer Genome Consortium Data Portal. Nat Biotechnol 2019; 37:367–369
23. Chakravarty D, Gao J, Phillips S, et al. OncoKB: A Precision Oncology Knowledge Base. JCO Precis Oncol 2017; 1–16
24. Berman HM, Westbrook J, Feng Z, et al. The Protein Data Bank. Nucleic Acids Res 2000; 28:235–242
25. Jumper J, Evans R, Pritzel A, et al. Highly accurate protein structure prediction with AlphaFold. Nature 2021; 596:583–589
26. Varadi M, Anyango S, Deshpande M, et al. AlphaFold Protein Structure Database: massively expanding the structural coverage of protein-sequence space with high-accuracy models. Nucleic Acids Res 2022; 50:D439–D444
27. Lin Z, Akin H, Rao R, et al. Evolutionary-scale prediction of atomic level protein structure with a language model. biorxiv 2021;
28. Martínez-Jiménez F, Muiños F, Sentís I, et al. A compendium of mutational cancer driver genes. Nat Rev Cancer 2020; 20:555–572
29. Bailey MH, Tokheim C, Porta-Pardo E, et al. Comprehensive Characterization of Cancer Driver Genes and Mutations. Cell 2018; 173:371-385.e18
30. Sondka Z, Bamford S, Cole CG, et al. The COSMIC Cancer Gene Census: describing genetic dysfunction across all human cancers. Nat Rev Cancer 2018; 18:696–705
31. Zhang P, Itan Y. Biological network approaches and applications in rare disease studies. Genes (Basel) 2019; 10:
32. Pham VVH, Liu L, Bracken CP, et al. CBNA: A control theory based method for identifying coding and non-coding cancer drivers. PLoS Comput Biol 2019; 15:e1007538
33. Wei PJ, Wu FX, Xia J, et al. Prioritizing Cancer Genes Based on an Improved Random Walk Method. Front Genet 2020; 11:

34. Akhavan-Safar M, Teimourpour B, Kargari M. GenHITS: A network science approach to driver gene detection in human regulatory network using gene's influence evaluation. J Biomed Inform 2021; 114:

35. Akhavan-Safar M, Teimourpour B. KatzDriver: A network based method to cancer causal genes discovery in gene regulatory network. BioSystems 2021; 201:

36. Rahimi M, Teimourpour B, Marashi SA. Cancer driver gene discovery in transcriptional regulatory networks using influence maximization approach. Comput Biol Med 2019; 114:

37. Pham VVH, Liu L, Bracken CP, et al. DriverGroup: A novel method for identifying driver gene groups. Bioinformatics 2020; 36:I583–I591

38. Elliott K, Larsson E. Non-coding driver mutations in human cancer. Nat Rev Cancer 2021; 21:500–509

39. Champion M, Brennan K, Croonenborghs T, et al. Module Analysis Captures Pancancer Genetically and Epigenetically Deregulated Cancer Driver Genes for Smoking and Antiviral Response. EBioMedicine 2018; 27:156–166

40. Wei PJ, Zhang D, Li HT, et al. DriverFinder: A gene length-based network method to identify cancer driver genes. Complexity 2017; 2017:

41. Dinstag G, Shamir R. PRODIGY: Personalized prioritization of driver genes. Bioinformatics 2020; 36:1831–1839

42. Wei PJ, Zhang D, Xia J, et al. LNDriver: Identifying driver genes by integrating mutation and expression data based on gene-gene interaction network. BMC Bioinformatics 2016; 17:221–230

43. Song J, Peng W, Wang F. A random walk-based method to identify driver genes by integrating the subcellular localization and variation frequency into bipartite graph. BMC Bioinformatics 2019; 20:

44. Ahmed R, Baali I, Erten C, et al. MEXCOwalk: Mutual exclusion and coverage based random walk to identify cancer modules. Bioinformatics 2020; 36:872–879

45. Peng W, Yi S, Dai W, et al. Identifying and ranking potential cancer drivers using representation learning on attributed network. Methods 2021; 192:13–24

46. Gumpinger AC, Lage K, Horn H, et al. Prediction of cancer driver genes through network-based moment propagation of mutation scores. Bioinformatics 2020; 36:I508–I515

47. Han Y, Yang J, Qian X, et al. DriverML: A machine learning algorithm for identifying driver genes in cancer sequencing studies. Nucleic Acids Res 2019; 47:e45

48. Guo WF, Zhang SW, Zeng T, et al. Network control principles for identifying personalized driver genes in cancer. Brief Bioinform 2020; 21:1641–1662

49. Gonzalez-Perez A, Lopez-Bigas N. Functional impact bias reveals cancer drivers. Nucleic Acids Res 2012; 40:e169

50. Nulsen J, Misetic H, Yau C, et al. Pan-cancer detection of driver genes at the single-patient resolution. Genome Med 2021; 13:

51. Ülgen E, Sezerman OU. driveR: a novel method for prioritizing cancer driver genes using somatic genomics data. BMC Bioinformatics 2021; 22:

52. Wang K, Li M, Hakonarson H. ANNOVAR: Functional annotation of genetic variants from high-throughput sequencing data. Nucleic Acids Res 2010; 38:

53. Yang H, Robinson PN, Wang K. Phenolyzer: Phenotype-based prioritization of candidate genes for human diseases. Nat Methods 2015; 12:841–843

54. Kanehisa M, Furumichi M, Tanabe M, et al. KEGG: new perspectives on genomes, pathways, diseases and drugs. Nucleic Acids Res 2017; 45:D353–D361

55. Peng W, Tang Q, Dai W, et al. Improving cancer driver gene identification using multi-Task learning on graph convolutional network. Brief Bioinform 2022; 23:1–12

56. Gu H, Xu X, Qin P, et al. FI-Net: Identification of Cancer Driver Genes by Using Functional Impact Prediction Neural Network. Front Genet 2020; 11:
57. Luo P, Ding Y, Lei X, et al. DeepDriver: Predicting cancer driver genes based on somatic mutations using deep convolutional neural networks. Front Genet 2019; 10:
58. Zeng Z, Mao C, Vo A, et al. Deep learning for cancer type classification and driver gene identification. BMC Bioinformatics 2021; 22:
59. Collier O, Stoven V, Vert JP. LOTUS: A single- And multitask machine learning algorithm for the prediction of cancer driver genes. PLoS Comput Biol 2019; 15:e1007381
60. Shi X, Teng H, Shi L, et al. Comprehensive evaluation of computational methods for predicting cancer driver genes. Brief Bioinform 2022; 23:1–14
61. Bertrand D, Chng KR, Sherbaf FG, et al. Patient-specific driver gene prediction and risk assessment through integrated network analysis of cancer omics profiles. Nucleic Acids Res 2015; 43:e44
62. Zhang W, Wang SL. A Novel Method for Identifying the Potential Cancer Driver Genes Based on Molecular Data Integration. Biochem Genet 2020; 58:16–39
63. Zhang D, Bin Y. DriverSubNet: A Novel Algorithm for Identifying Cancer Driver Genes by Subnetwork Enrichment Analysis. Front Genet 2021; 11:
64. Pan H, Renaud L, Chaligne R, et al. Discovery of candidate dna methylation cancer driver genes. Cancer Discov 2021; 11:2266–2281
65. Li A, Chapuy B, Varelas X, et al. Identification of candidate cancer drivers by integrative Epi-DNA and Gene Expression (iEDGE) data analysis. Sci Rep 2019; 9:
66. Lu X, Wang X, Ding L, et al. FrDriver: A Functional Region Driver Identification for Protein Sequence. IEEE/ACM Trans Comput Biol Bioinform 2021; 18:1773–1783
67. Arnedo-Pac C, Mularoni L, Muiños F, et al. OncodriveCLUSTL: A sequence-based clustering method to identify cancer drivers. Bioinformatics 2019; 35:4788–4790
68. Zhu H, Uusküla-Reimand L, Isaev K, et al. Candidate Cancer Driver Mutations in Distal Regulatory Elements and Long-Range Chromatin Interaction Networks. Mol Cell 2020; 77:1307-1321.e10
69. Bokhari Y, Alhareeri A, Arodz T. QuaDMutNetEx: A method for detecting cancer driver genes with low mutation frequency. BMC Bioinformatics 2020; 21:
70. Hou Y, Gao B, Li G, et al. MaxMIF: A New Method for Identifying Cancer Driver Genes through Effective Data Integration. Advanced Science 2018; 5:
71. Zapata L, Susak H, Drechsel O, et al. Signatures of positive selection reveal a universal role of chromatin modifiers as cancer driver genes. Sci Rep 2017; 7:
72. Korthauer KD, Kendziorski C. MADGiC: A model-based approach for identifying driver genes in cancer. Bioinformatics 2015; 31:1526–1535
73. Datta N, Chakraborty S, Basu M, et al. Tumor Suppressors Having Oncogenic Functions: The Double Agents. Cells 2020; 10:1–26
74. Stepanenko AA, Vassetzky YS, Kavsan VM. Antagonistic functional duality of cancer genes. Gene 2013; 529:199–207
75. Croce CM. Oncogenes and cancer. N Engl J Med 2008; 358:502–511
76. Wang LH, Wu CF, Rajasekaran N, et al. Loss of Tumor Suppressor Gene Function in Human Cancer: An Overview. Cellular Physiology and Biochemistry 2018; 51:2647–2693
77. Shen L, Shi Q, Wang W. Double agents: Genes with both oncogenic and tumor-suppressor functions. Oncogenesis 2018; 7:
78. Chandrashekar P, Ahmadinejad N, Wang J, et al. Somatic selection distinguishes oncogenes and tumor suppressor genes. Bioinformatics 2020; 36:1712–1717

79. Tokheim CJ, Papadopoulos N, Kinzler KW, et al. Evaluating the evaluation of cancer driver genes. Proc Natl Acad Sci U S A 2016; 113:14330–14335
80. Lyu J, Li JJ, Su J, et al. DORGE: Discovery of Oncogenes and tumoR suppressor genes using Genetic and Epigenetic features. Sci Adv 2020; 6:
81. Colaprico A, Olsen C, Bailey MH, et al. Interpreting pathways to discover cancer driver genes with Moonlight. Nat Commun 2020; 11:69
82. Saksager A, Nourbakhsh M, Tom N, et al. An Automatized Workflow to Study Mechanistic Indicators for Driver Gene Prediction with Moonlight. bioRxiv 2022;
83. Rajendran BK, Deng C-X. Characterization of potential driver mutations involved in human breast cancer by computational approaches. Oncotarget 2017; 8:50252–50272
84. Repana D, Nulsen J, Dressler L, et al. The Network of Cancer Genes (NCG): a comprehensive catalogue of known and candidate cancer genes from cancer sequencing screens. Genome Biol 2019; 20:1
85. Parvandeh S, Donehower LA, Katsonis P, et al. EPIMUTESTR: a nearest neighbor machine learning approach to predict cancer driver genes from the evolutionary action of coding variants. Nucleic Acids Res 2022; 50:e70–e70
86. Darbyshire M, du Toit Z, Rogers MF, et al. Estimating the Frequency of Single Point Driver Mutations across Common Solid Tumours. Sci Rep 2019; 9:
87. Martincorena I, Raine KM, Gerstung M, et al. Universal Patterns of Selection in Cancer and Somatic Tissues. Cell 2017; 171:1029-1041.e21
88. Dong C, Wei P, Jian X, et al. Comparison and integration of deleteriousness prediction methods for nonsynonymous SNVs in whole exome sequencing studies. Hum Mol Genet 2015; 24:2125–2137
89. Wang H, Wang T, Zhao X, et al. AI-Driver: an ensemble method for identifying driver mutations in personal cancer genomes. NAR Genom Bioinform 2020; 2:lqaa084
90. Davydov E V., Goode DL, Sirota M, et al. Identifying a High Fraction of the Human Genome to be under Selective Constraint Using GERP++. PLoS Comput Biol 2010; 6:e1001025
91. Jagadeesh KA, Wenger AM, Berger MJ, et al. M-CAP eliminates a majority of variants of uncertain significance in clinical exomes at high sensitivity. Nat Genet 2016; 48:1581–1586
92. Rogers MF, Gaunt TR, Campbell C. CScape-somatic: Distinguishing driver and passenger point mutations in the cancer genome. Bioinformatics 2020; 36:3637–3644
93. Mao Y, Chen H, Liang H, et al. CanDrA: Cancer-Specific Driver Missense Mutation Annotation with Optimized Features. PLoS One 2013; 8:e77945
94. Choi Y, Chan AP. PROVEAN web server: A tool to predict the functional effect of amino acid substitutions and indels. Bioinformatics 2015; 31:2745–2747
95. Kumar P, Henikoff S, Ng PC. Predicting the effects of coding non-synonymous variants on protein function using the SIFT algorithm. Nat Protoc 2009; 4:1073–1082
96. Vaser R, Adusumalli S, Leng SN, et al. SIFT missense predictions for genomes. Nat Protoc 2016; 11:1–9
97. Kircher M, Witten DM, Jain P, et al. A general framework for estimating the relative pathogenicity of human genetic variants. Nat Genet 2014; 46:310–315
98. Brown A-L, Li M, Goncearenco A, et al. Finding driver mutations in cancer: Elucidating the role of background mutational processes. PLoS Comput Biol 2019; 15:e1006981
99. Laine E, Karami Y, Carbone A. GEMME: A Simple and Fast Global Epistatic Model Predicting Mutational Effects. Mol Biol Evol 2019; 36:2604–2619
100. Munro D, Singh M. DeMaSk: A deep mutational scanning substitution matrix and its use for variant impact prediction. Bioinformatics 2020; 36:5322–5329

101. Ioannidis NM, Rothstein JH, Pejaver V, et al. REVEL: An Ensemble Method for Predicting the Pathogenicity of Rare Missense Variants. Am J Hum Genet 2016; 99:877–885
102. Ionita-Laza I, Mccallum K, Xu B, et al. A spectral approach integrating functional genomic annotations for coding and noncoding variants. Nat Genet 2016; 48:214–220
103. Pollard KS, Hubisz MJ, Rosenbloom KR, et al. Detection of nonneutral substitution rates on mammalian phylogenies. Genome Res 2010; 20:110–121
104. Li J, Zhao T, Zhang Y, et al. Performance evaluation of pathogenicity-computation methods for missense variants. Nucleic Acids Res 2018; 46:7793–7804
105. Tokheim C, Bhattacharya R, Niknafs N, et al. Exome-scale discovery of hotspot mutation regions in human cancer using 3D protein structure. Cancer Res 2016; 76:3719–3731
106. Raimondi D, Tanyalcin I, FertCrossed JSD, et al. DEOGEN2: Prediction and interactive visualization of single amino acid variant deleteriousness in human proteins. Nucleic Acids Res 2017; 45:W201–W206
107. Rentzsch P, Witten D, Cooper GM, et al. CADD: Predicting the deleteriousness of variants throughout the human genome. Nucleic Acids Res 2019; 47:D886–D894
108. Shihab HA, Gough J, Cooper DN, et al. Predicting the functional consequences of cancer-associated amino acid substitutions. Bioinformatics 2013; 29:1504–1510
109. Rogers MF, Shihab HA, Mort M, et al. FATHMM-XF: Accurate prediction of pathogenic point mutations via extended features. Bioinformatics 2018; 34:511–513
110. Yue Z, Chu X, Xia J. PredCID: Prediction of driver frameshift indels in human cancer. Brief Bioinform 2021; 22:
111. Pagel KA, Pejaver V, Lin GN, et al. When loss-of-function is loss of function: Assessing mutational signatures and impact of loss-of-function genetic variants. Bioinformatics 2017; 33:i389–i398
112. Pagel KA, Antaki D, Lian A, et al. Pathogenicity and functional impact of non-frameshifting insertion/deletion variation in the human genome. PLoS Comput Biol 2019; 15:e1007112
113. Sundaram L, Gao H, Padigepati SR, et al. Predicting the clinical impact of human mutation with deep neural networks. Nat Genet 2018; 50:1161–1170
114. Song K, Li Q, Gao W, et al. AlloDriver: A method for the identification and analysis of cancer driver targets. Nucleic Acids Res 2019; 47:W315–W321
115. Lu Q, Hu Y, Sun J, et al. A Statistical Framework to Predict Functional Non-Coding Regions in the Human Genome Through Integrated Analysis of Annotation Data. Sci Rep 2015; 5:10576
116. Frazer J, Notin P, Dias M, et al. Disease variant prediction with deep generative models of evolutionary data. Nature 2021; 599:91–95
117. Livesey BJ, Marsh JA. Using deep mutational scanning to benchmark variant effect predictors and identify disease mutations. Mol Syst Biol 2020; 16:e9380
118. Chen H, Li J, Wang Y, et al. Comprehensive assessment of computational algorithms in predicting cancer driver mutations. Genome Biol 2020; 21:43
119. Rogers MF, Gaunt TR, Campbell C. Prediction of driver variants in the cancer genome via machine learning methodologies. Brief Bioinform 2021; 22:bbaa250
120. Cheng F, Zhao J, Wang Y, et al. Comprehensive characterization of protein–protein interactions perturbed by disease mutations. Nat Genet 2021; 53:342–353
121. Kryshtafovych A, Schwede T, Topf M, et al. Critical assessment of methods of protein structure prediction (CASP)—Round XIV. Proteins: Structure, Function and Bioinformatics 2021; 89:1607–1617

122. Kumar S, Clarke D, Gerstein MB. Leveraging protein dynamics to identify cancer mutational hotspots using 3D structures. Proc Natl Acad Sci U S A 2019; 116:18962–18970
123. Porta-Pardo E, Garcia-Alonso L, Hrabe T, et al. A Pan-Cancer Catalogue of Cancer Driver Protein Interaction Interfaces. PLoS Comput Biol 2015; 11:e1004518
124. Sora V, Tiberti M, Robbani SM, et al. PyInteraph2 and PyInKnife2 to analyze networks in protein structural ensembles. bioRxiv 2020;
125. Ng PKS, Li J, Jeong KJ, et al. Systematic Functional Annotation of Somatic Mutations in Cancer. Cancer Cell 2018; 33:450-462.e10
126. Zhang H, Xu MS, Fan X, et al. Predicting functional effect of missense variants using graph attention neural networks. Nat Mach Intell 2022; 4:1017–1028
127. Meyer MJ, Lapcevic R, Romero AE, et al. mutation3D: Cancer Gene Prediction Through Atomic Clustering of Coding Variants in the Structural Proteome. Hum Mutat 2016; 37:447–456
128. Niu B, Scott AD, Sengupta S, et al. Protein-structure-guided discovery of functional mutations across 19 cancer types. Nat Genet 2016; 48:827–837
129. Kamburov A, Lawrence MS, Polak P, et al. Comprehensive assessment of cancer missense mutation clustering in protein structures. Proc Natl Acad Sci U S A 2015; 112:E5486–E5495
130. Landrum MJ, Lee JM, Riley GR, et al. ClinVar: Public archive of relationships among sequence variation and human phenotype. Nucleic Acids Res 2014; 42:D980-5
131. Gao J, Chang MT, Johnsen HC, et al. 3D clusters of somatic mutations in cancer reveal numerous rare mutations as functional targets. Genome Med 2017; 9:4
132. Sivley RM, Dou X, Meiler J, et al. Comprehensive Analysis of Constraint on the Spatial Distribution of Missense Variants in Human Protein Structures. Am J Hum Genet 2018; 102:415–426
133. Gress A, Ramensky V, Büch J, et al. StructMAn: annotation of single-nucleotide polymorphisms in the structural context. Nucleic Acids Res 2016; 44:W463–W468
134. Hicks M, Bartha I, di Iulio J, et al. Functional characterization of 3D protein structures informed by human genetic diversity. Proceedings of the National Academy of Sciences 2019; 116:8960–8965
135. Tiberti M, Terkelsen T, Degn K, et al. MutateX: an automated pipeline for in silico saturation mutagenesis of protein structures and structural ensembles. Brief Bioinform 2022; 23:bbac074
136. Valanciute A, Nygaard L, Zschach H, et al. Accurate protein stability predictions from homology models. Comput Struct Biotechnol J 2023; 21:66–73
137. Akdel M, Pires DE V., Pardo EP, et al. A structural biology community assessment of AlphaFold2 applications. Nat Struct Mol Biol 2022; 29:1056–1067
138. Blaabjerg LM, Kassem MM, Good LL, et al. Rapid protein stability prediction using deep learning representations. bioRxiv 2022;
139. Iqbal S, Hoksza D, Pérez-Palma E, et al. MISCAST: MIssense variant to protein structure analysis web suite. Nucleic Acids Res 2021; 48:W132–W139
140. Nygaard M, Terkelsen T, Olsen AV, et al. The mutational landscape of the oncogenic MZF1 SCAN domain in cancer. Front Mol Biosci 2016; 3:1–18
141. Kumar M, Papaleo E. A pan-cancer assessment of alterations of the kinase domain of ULK1, an upstream regulator of autophagy. Sci Rep 2020; 10:14874
142. König SM, Rissler V, Terkelsen T, et al. Alterations of the interactome of Bcl-2 proteins in breast cancer at the transcriptional, mutational and structural level. PLoS Comput Biol 2019; 15:e1007485

143. Degn K, Beltrame L, Dahl Hede F, et al. Cancer-related Mutations with Local or Long-range Effects on an Allosteric Loop of p53. J Mol Biol 2022; 434:
144. Arnaudi M, Beltrame L, Degn K, et al. MAVISp: Multi-layered Assessment of VarIants by Structure for proteins. bioRxiv 2022; 1–12
145. Tate JG, Bamford S, Jubb HC, et al. COSMIC: The Catalogue Of Somatic Mutations In Cancer. Nucleic Acids Res 2019; 47:D941–D947
146. Gao J, Aksoy BA, Dogrusoz U, et al. Integrative analysis of complex cancer genomics and clinical profiles using the cBioPortal. Sci Signal 2013; 6:pl1
147. Tiberti M, Di Leo L, Vistesen MV, et al. The Cancermuts software package for the prioritization of missense cancer variants: a case study of AMBRA1 in melanoma. Cell Death Dis 2022; 13:872
148. Høie MH, Cagiada M, Beck Frederiksen AH, et al. Predicting and interpreting large-scale mutagenesis data using analyses of protein stability and conservation. Cell Rep 2022; 38:110207
149. Cagiada M, Johansson KE, Valanciute A, et al. Understanding the Origins of Loss of Protein Function by Analyzing the Effects of Thousands of Variants on Activity and Abundance. Mol Biol Evol 2021; 38:3235–3246
150. Hyman DM, Taylor BS, Baselga J. Implementing Genome-Driven Oncology. Cell 2017; 168:584–599

**Figure 1: Illustration of the Concept of Driver Genes, Driver Mutations, and Structural Impact.** (A) Cancer involves dynamic changes in the genome caused by genomic alterations such as mutations, epigenetic changes or chromosomal rearrangements. These genomic alterations occur in driver genes which are divided into oncogenes, tumor suppressors, and dual role genes. Oncogenes normally promote cell growth, whereas tumor suppressors normally limit cell growth. The dual role genes exhibit both tumor suppressive and oncogenic behavior depending on the cellular context. The genomic alterations lead to gain of function of oncogenes and loss of function of tumor suppressors which leads to uncontrolled cell growth and cancer. (B) Mutations can be categorized as passengers and drivers. Passengers are characterized by the absence of pathogenic impact. This can be due to their placement away from functional sites of the coding region or regulatory elements in the non-coding regions or the nature of the resulting amino acid substitution. (C) Genes are translated into proteins, and potential alterations including mutations are thereby also expressed. Understanding how these are expressed can give a mechanistic understanding as to why a particular alteration may give the cells a growth advantage. The top panel illustrates the healthy process of interaction, while the bottom panel illustrates some of the structural expressions of driver mutations.

**Figure 2: Overview of Driver Gene Prediction Tools.** All driver gene prediction tools discussed in this review are listed in the middle. These tools are divided into four main categories based on the underlying computational architecture. These four categories are subcategorized and to the right, each is listed. \*\*driver gene prediction tools that predict tumor suppressors, oncogenes, and dual role genes. \*driver gene prediction tools that predict tumor suppressors and oncogenes.

**Figure 3: The Moonlight Framework for Driver Gene Prediction.** Moonlight uses a set of differentially expressed genes (DEGs) as input. First, a functional enrichment analysis is carried out to find which of Moonlight's 101 biological processes (BPs) are overrepresented among the DEGs. Then, a gene regulatory network analysis models how the DEGs are connected with each other through mutual information. Following this step, Moonlight diverges into an expert-based and a machine learning approach. In the next step, an upstream regulatory analysis, the expert-based approach examines the effect of DEGs on user-selected biological processes whereas the machine learning approach examines this on all of Moonlight's biological processes. Subsequently, putative tumor suppressors and oncogenes collectively called oncogenic mediators are predicted through a pattern recognition analysis using either patterns (the expert-based approach) or a random forest classifier (the machine learning approach). Finally, a driver mutation analysis analyzes mutations in the cancer patient cohort and categorizes these into drivers and passengers. Those oncogenic mediators containing at least one driver mutation are retained as driver genes.

**Figure 4: Overview of Driver Mutation Prediction Tools.** All driver mutation prediction tools discussed in this review are listed in the middle. These tools are divided into three main categories based on the underlying computational architecture. These three categories are subcategorized and to the right, each is listed.

**Figure 5: Overview of Structural Analysis Tools**. All driver structural analysis tools discussed in this review are listed in the middle. These tools are divided into four categories based on the underlying computational architecture. To the right, each tool is listed.

**Figure 6: MAVISp Framework for Structural Analysis.** Multi-layered Assessment of Varlants by Structure for proteins (MAVISp) is a module dependent framework to study structural alterations. The curators of MAVISp take mutations as an input and study these using the applicable modules starting with the structure selection and potentially ensemble generation. The mutations are modeled in the structure and analyzed for their impact on stability, local interactions, long-range effects, post translational modifications and if an ensemble was generated, functional dynamics. All of these analyses provide information for variant assessment. *Modules are part of the ensemble-mode of MAVISp.

**Figure 7: Current and Suggested Future Workflow.** (A) Current research within driver gene prediction, driver mutation prediction, and structural assessment of mutations is characterized by tools within these fields working in silos. Such an approach risks overlooking potential synergies. (B) To overcome some of the challenges by this approach, a suggested protocol for the future development of these fields is illustrated. This protocol includes the collective study of the driver gene prediction, driver mutation, and structural assessment of mutation fields, with each field serving as the input to the next, creating a funnel approach. Such a novel workflow would include initial driver gene prediction, selection of driver genes, driver mutation prediction in the predicted driver genes, and finally, choice of mutations to be studied structurally to assess the impact of these mutations on protein function and stability. This will result in a set of damaged proteins for further exploration.

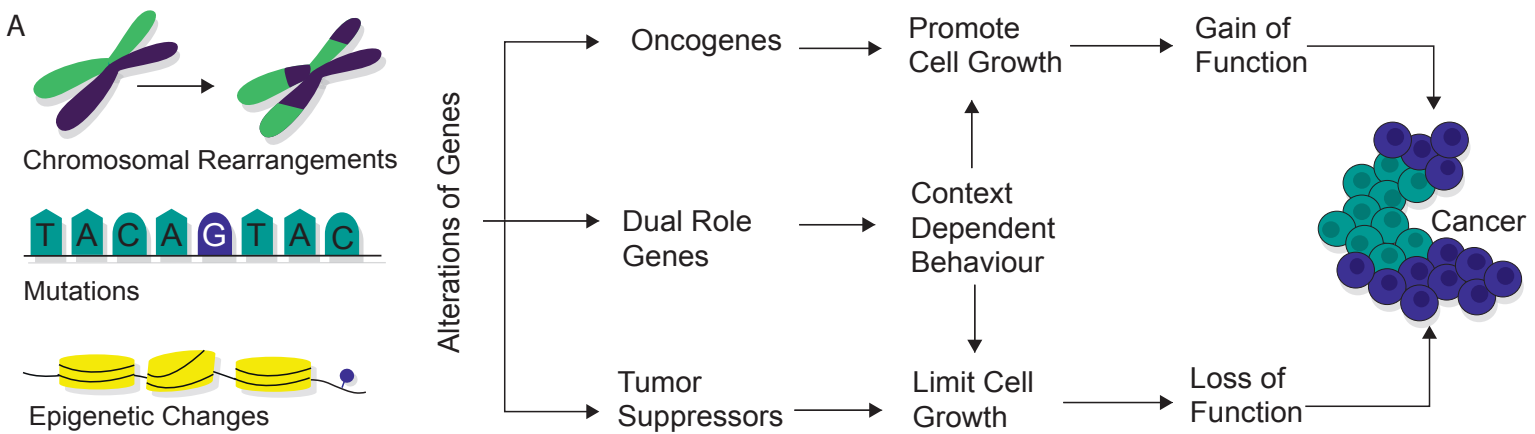

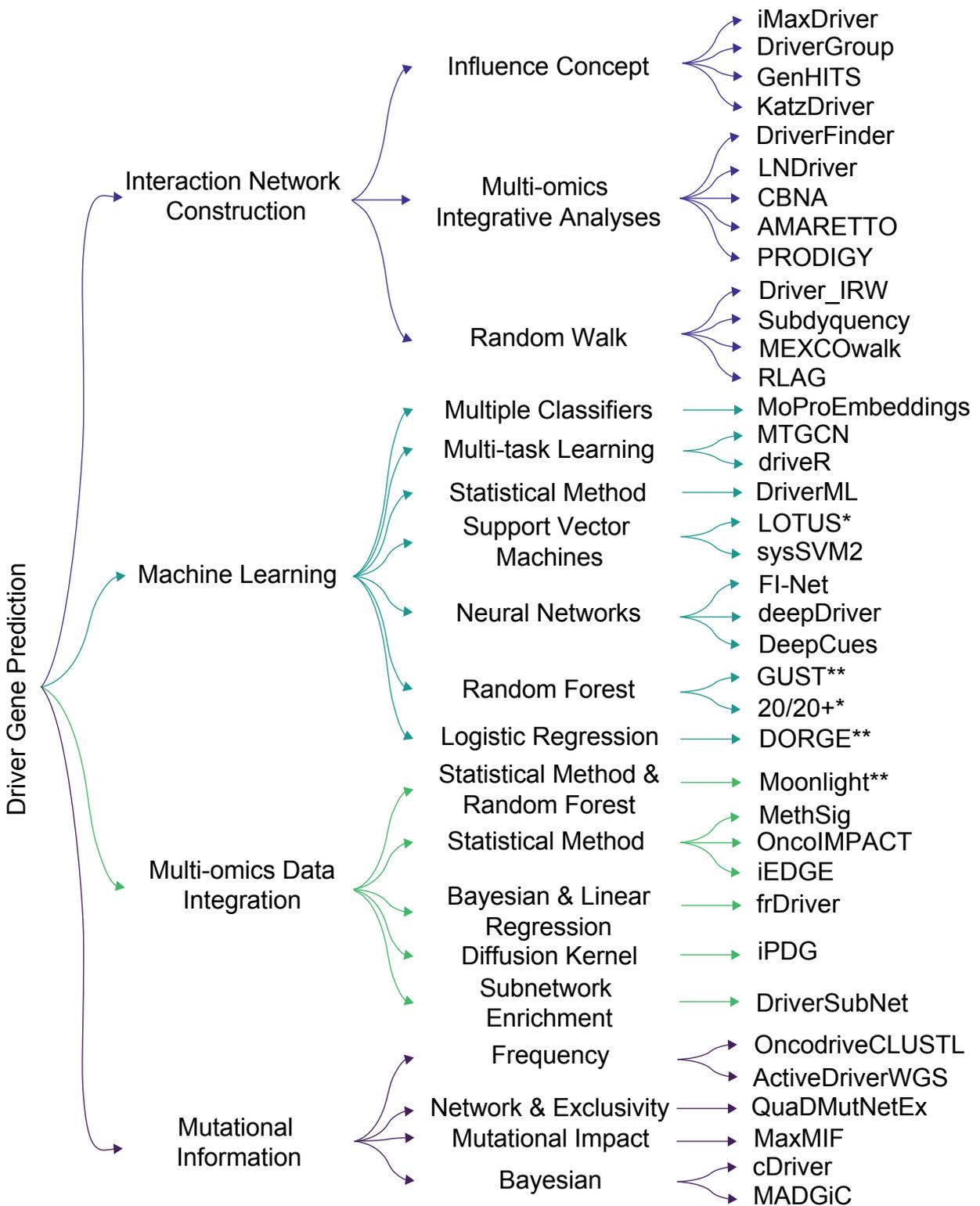

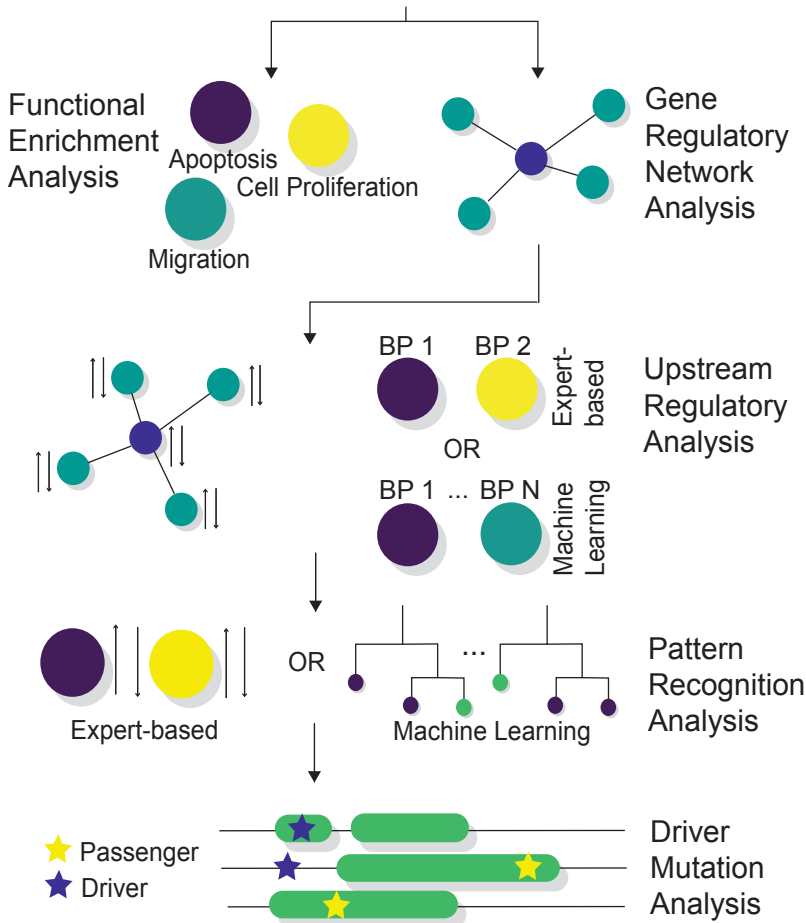

```
Driver Mutation Prediction
├── Statistical Methods
│   └── Evolutionary Conservation Score
│       ├── DriverML
│       ├── SIFT
│       ├── PROVEAN
│       ├── DeMaSk
│       └── GEMME
├── Supervised Machine Learning Methods
│   ├── Neural Network
│   │   ├── MutPred-LOF
│   │   └── MutPredInDel
│   ├── Gradient Boosting
│   │   ├── AI-Driver
│   │   ├── BoostDM
│   │   ├── PredCID
│   │   ├── Cscape
│   │   └── Cscape-Somatic
│   ├── Hidden Markov Models
│   │   └── FATHMM
│   ├── Random Forest
│   │   ├── REVEL
│   │   ├── DENOGEN2
│   │   ├── CHASMplus
│   │   └── M-CAP
│   └── Support Vector Machines & Logistic Regression
│       └── CADD
└── Unsupervised Machine Learning Methods
    ├── Combination Tools
    │   ├── GenoCanyon
    │   ├── Eigen
    │   └── Allodriver
    ├── Deep Neural Networks
    │   └── PrimateAI
    └── Generative Modeling
        └── EVE
```

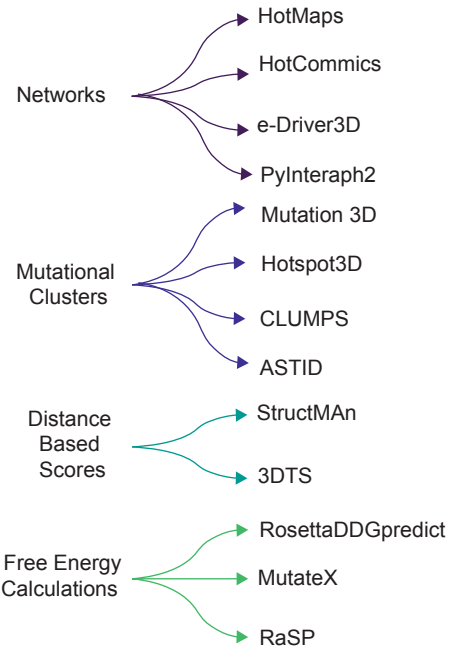

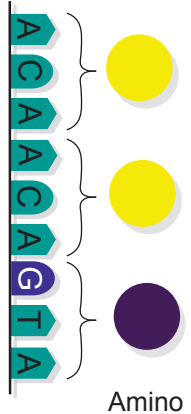
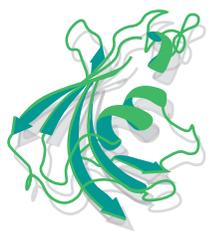
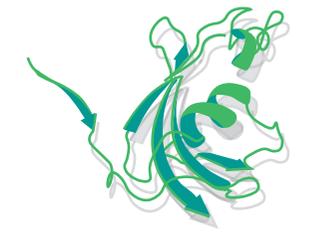
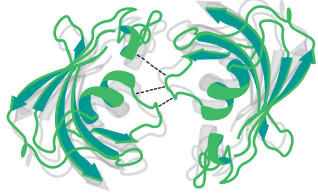
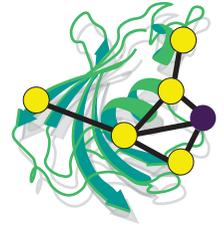
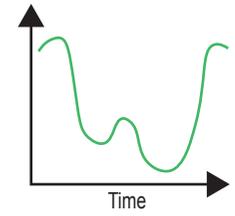
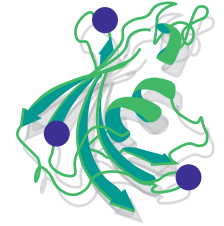
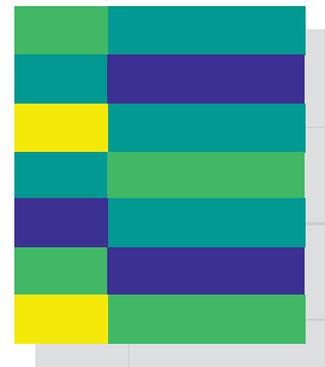

Mutation of Interest — Amino Acid — Structure Selection & Ensemble Generation*

MAVISp Analysis Modules: Stability; Interactome & Local Interactions; Long-range Effects; Functional Dynamics*; Post Translational Modifications

Variant Annotation & Classification

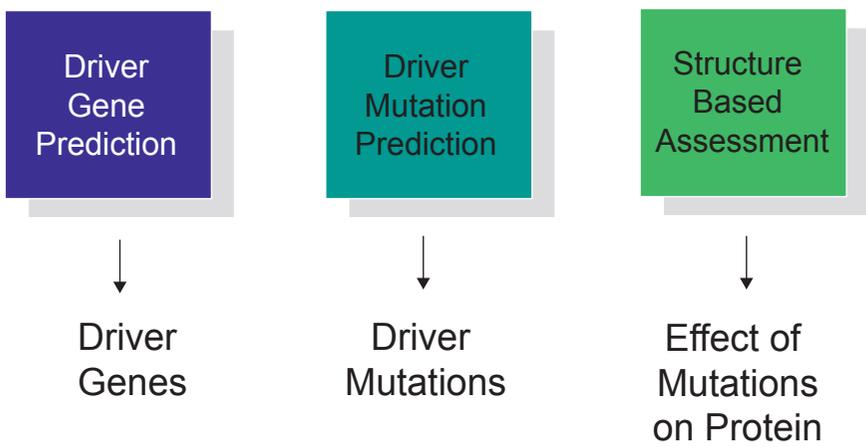
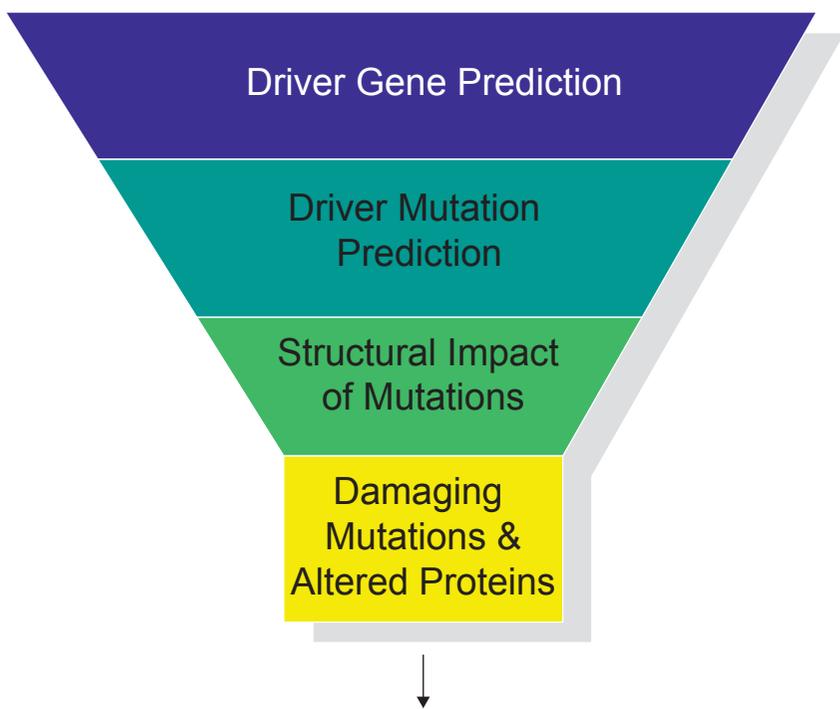

# Supplementary materials

**Table S1: Datasets and Databases - Large Non-annotated Databases**

| Database name | Datatype | Comments | Access to database |
|---|---|---|---|
| National Cancer Institute's (NCI's) Genomic Data Commons (GDC) (10.1038/s41588-021-00791-5; 10.1038/s41467-021-21254-9; 10.1056/NEJMp1607591) | Genomic, transcriptomic, epigenetic, proteomics, and clinical data. | GDC is a data repository, data sharing platform, and knowledge network, aiming overall at promoting cancer precision medicine and supporting a standardized format across various cancer genome programs such as TCGA, TARGET, GENIE, CPTAC, and CCLE, FMI, MMRF, CMI, CGCI, and HCMI. | https://gdc.cancer.gov/ |
| NCI's Proteomic Data Commons (PDC) (https://doi.org/10.1158/1538-7445.AM2020-LB-242) | Proteomic data. | Contains a range of cancer proteomic data sets stored in a unified repository. PDC contains around 40 datasets from more than 12 cancer types generated by various large-scale cancer research studies. | https://pdc.cancer.gov/pdc/ |
| the International Cancer Genome Consortium (ICGC) (10.1038/s41587-019-0055-9) | Genomic, transcriptomic, epigenomic, proteomics, and clinical data. | Annotated database with somatic information from a range of cancer projects. Moreover, it is a data portal containing data from 84 cancer projects. | https://dcc.icgc.org/ |
| cBioPortal (10.1158/2159-8290.CD-12-0095; 10.1126/scisignal.2004088) | Genomic, transcriptomic, epigenomic, and clinical data. | Hosts datasets from several large-scale cancer research initiatives including information about the variations such as mRNA expression, mutations, copy number, DNA methylation, clinical outcome, and post translational modifications. cBioPortal provides data from more than 5000 tumor samples from 20 cancer studies. | http://cbioportal.org |
| St. Jude Cloud (10.1158/2159-8290.CD-20-1230) | Genomic and transcriptomic data. | A platform where genomic data from whole genomes, whole exomes and transcriptomes from more than 10,000 pediatric cancer patients can be accessed, analyzed and visualized. | https://www.stjude.cloud |

**Table S2: Datasets and Databases - Annotated Gene Databases**

| Database | Curration | Datatype | Comments | Access to |
|---|---|---|---|---|

| name | | | | database/dataset |
|---|---|---|---|---|
| Cancer Gene Census (CGC) (doi.org/10.1038/s41568-018-0060-1) | Manual literature curation | Hallmarks, coding mutation, tissue and cell type, and tumor type. | A resource available in COSMIC which contains a description of genes that drive human cancer which is used as a standard in cancer genetics. The 2018 Cancer Gene Census describes the effect of 719 cancer driver genes. The genes must be reported in at least two different studies and have experiential evidence. | https://cancer.sanger.ac.uk/census |
| Catalogue Of Somatic Mutations In Cancer (10.1093/nar/gky1015; 10.1093/nar/gkw1121; 10.1093/nar/gku1075) | Manual curation and a semi-automated process | Coding mutations, non-coding mutations, gene fusions, copy number variants and drug-resistance mutations. | Currently, the most comprehensive database storing somatic mutations across human cancers. Today, COSMIC includes all human genes and around 6 million coding mutations. COSMIC also includes further resources: COSMIC Cell Lines Project which aims at characterizing genomics of cancer cell lines and COSMIC-3D which connects sequence-level mutation and protein structural data. | https://cancer.sanger.ac.uk |
| the Network of Cancer Genes (NCG) (10.1186/s13059-018-1612-0) | Manual, collected from literature and from Cancer Gene Census and list of driver genes by Vogelstein et al. 2013 | Classification of drivers. Germline variations e.g. SNV, indels, structural variations. Gene and protein expression, protein and miRNA interaction, evolutionary conservation, gene duplication, gene function, drug target/biomarkers. | A gene database of 2372 cancer driver genes. The aim is to comprehensively describe genes with a vast amount of annotations such as evolutionary origin, gene and protein expression, gene essentiality and interactions. | http://ncg.kcl.ac.uk/ |
| DriverDBv3 (10.1093/nar/gkz964) | Computational curation using published bioinformatics algorithms/tools | Incorporates somatic mutation, RNA expression, miRNA expression, methylation, copy number variation, and clinical data. | A multi-omics cancer driver gene database. They collected data from TCGA and obtained information about cancer genes from the Cancer Gene Census in COSMIC and the Network of Cancer Genes. They used various bioinformatics tools to analyze molecular features in driver genes, covering mutation, methylation, copy number variation, miRNA, and survival events. | http://ngs.ym.edu.tw/driverdb |
| ONGene (10.1016/j.jgg.2016.12.004) | Literature mining followed by manual curation | Investigation of curated oncogenes including, literature data, gene expression profiles, mutation patterns, lncRNA interactome analysis, homologs, and information about biological pathways. | ONGene contains 803 human oncogenes, of which 698 are protein-coding and 105 are non-coding, connected to 8849 curated PubMed abstracts. | http://ongene.bioinfo-minzhao.org/ |
| TSGene (10.1093/nar/gkv1268; 10.1093/nar/gks937) | Literature mining followed by manual curation | Investigation of curated tumor suppressors including literature data, long non-coding tumor suppressor genes, curated tumor suppressor microRNAs, pan-cancer gene expression profiles, regulatory information, and information about biological pathways. | TSGene contains 1217 human tumor suppressor genes, of which 1018 are protein-coding and 199 are non-coding genes, curated from over 9000 articles. | https://bioinfo.uth.edu/TSGene/ |
| MSGene (10.1038/srep15478) | Literature mining followed by manual curation | Investigation of curated metastasis suppressor genes including literature data, biological pathways, gene expression profiles, regulation, mutations, interactions, and homologs. | Containing 194 experimentally verified metastasis suppressor genes which have been manually curated. Contains 194 experimentally verified metastasis suppressor genes and mapped to 1440 homologous genes from 17 model species, which have been done through manual curation. | http://MSGene.bioinfo-minzhao.org/ |
| dbEMT (10.1038/srep11459; 10.1016/j.jgg.2019.11.010) | Literature mining followed by manual curation | Investigation of EMT-related human genes including literature data, clinical relevant variants, gene expression profiles, biological pathways, co-expressed lncRNAs, regulation data, somatic mutations, homologs and interaction data. | Containing epithelial-mesenchymal transition related genesdbEMT2.0 contains 1184 human genes manually curated from more than 2665 PubMed abstracts (comprising 1011 protein-coding genes and 173 noncoding RNA genes). | http://dbemt.bioinfo-minzhao.org/ |
| LnCaNet (10.1093/bioinformatics/btw017) | Computational curation | Co-expression pairs of lncRNAs and cancer genes. | Containing co-expression pairs and interactions between lncRNAs and cancer genes. | http://lncanet.bioinfo-minzhao.org/ |
| Pedican (10.1038/srep11435) | Literature mining followed by manual curation | Annotations for each gene including literature data, involved biological pathways, interactions, mutations, gene expression, and regulation data. | Pedican focuses specifically on pediatric cancers and comprises 735 genes, 88 gene fusion and 24 chromosome events related to pediatric cancers, manually curated from 2245 PubMed abstracts. focusing on pediatric cancers. | http://pedican.bioinfo-minzhao.org/ |

**Table S3: Datasets and Databases - Annotated Mutational Databases**

| Database name | Curration | Datatype | Comments | Access to database/dataset |
|---|---|---|---|---|
| OncoVar (10.1093/nar/gkaa1033) | Computational data collection, structured manual filtering | Oncogenic driver mutations (coding and non-coding). | The data of OncoVar originates from 33 cancer types from TCGA and 18 cancer types from ICGC for somatic variants and gnomAD v2.1 for germline variants. The | https://oncovar.org/ |

| | | | data quality is optimized by exclusion of hypermutated tumors, passing of MC3 (Multicenter mutation-calling in multiple cancers network) filters and exclusion of tumors with inconsistent pathology. This subset of cleaned data was analyzed based on 23 predictive algorithms to assess the oncogenic nature of the mutations, genes, and pathways. | |
|---|---|---|---|---|
| IntOGen (10.1038/nmeth.2642; 10.1038/s41568-020-0290-x) | Computational ranking of genes. Tumor type is manually curated per cohort | Somatic mutations; single nucleotide variants and short indels (coding). | IntOGen is a framework for identifying driver genes in cohorts of cancers from somatic mutations. Each gene's mutations in each cohort is assessed by combining multiple computational methods and ranking them on the basis of bona fide cancer genes in COSMIC. The compendium contains 66 cancer types across 221 cohorts from other databases such as TCGA, ICGC, st. JUDE, TARGET, and CBIO. | https://www.intogen.org |
| Human Gene Mutation Database (HGMD) (10.1007/s00439-013-1358-4) | Manually literature based curation | Disease-causing mutations. | HGMD is a database with manually curated entries from literature. The database is aimed at clinicians and molecular geneticists as a reference framework for diagnostic testing and is maintained in collaboration with commercial partners. | http://www.hgmd.org |
| dbCMP (10.1093/bib/bby105) | Manually literature based curation | Passenger mutations. | The aim of the dataset is to provide a high quality negative dataset for benchmarking cancer mutational effects, especially for building novel computational predictive tools. The database contains 941 experimentally supported and 978 putative passenger mutations. | http://bioinfo.ahu.edu.cn:8080/dbCPM |
| ThermomutDB (10.1093/nar/gkaa925) | Manually literature based curation | Experimentally quantified mutations. | The mutations are quantified as their thermodynamic parameters, e.g. their impact on structural stability of the protein. | http://biosig.unimelb.edu.au/thermomutdb |
| OncoBase (10.1093/nar/gky1139) | Combination of multiple databases and computational assessment | Noncoding somatic mutations, epigenomics data, enhancer-promoter pairs, gene co-expression. | A collection of somatic mutations from TCGA, ICGC, COSMIC and ClinVar, and epigenomic data from ENCODE and Roadmap. The database has annotated somatic mutations in 68 cancer types from more than 120 cancer projects and explores the region's distal interactions with regulatory elements. It aims to have functional annotations of regulatory noncoding mutations and benchmark their effect in human carcinogenesis. | http://www.oncobase.biols.ac.cn/ |
| VARAdb (10.1093/nar/gkaa922) | Combination of multiple databases | Noncoding mutations. | Non-coding variants annotated with regulatory information, related genes, chromatin accessibility and chromatin interaction. | http://www.licpathway.net/VARAdb/ |
| dbCID (10.1093/bib/bby059) | Manual literature based curation | Insertions and deletions. | A database of insertions and deletions (indels) with literature evidence of being a cancer driver. For each indel there is information on gene, DNA, transcript and protein level. The data on gene level were obtained from CGC and from Pfam protein families data-base in protein level. The aim is to have a repository of human cancer indels to use in developing computational methods for identification of their effect in cancer. | http://bioinfo.ahu.edu.cn:8080/dbCID |
| CancerEnD (10.1016/j.ygeno.2020.04.028) | Combination of databases | Enhancers: Copy numbers, somatic mutations and survival. | Contains information about cancer associated enhancers. CancerEnD has annotated the enhancers in a cancer tissue-specific way in terms of expression, enhancer-gene interactions, somatic mutations, copy number variations and association with overall survival of cancer patients for 18 TCGA cancer types. The database is created through integrative analysis of RNA-seq gene expression, copy number variation, clinical data from TCGA,somatic | https://webs.iiitd.edu.in/raghava/cancerend/ |

| | | | mutation data from COSMIC, and enhancers from FANTOM. | |
|---|---|---|---|---|
| CanDL (10.1016/j.jmoldx.2015.05.002) | Manual curation based on revision of literature | Driver mutations. | Through literature reviews, they found variants that have been functionally characterized in vitro or in vivo as driver mutations. Chromosome location, all possible nucleotide positions for each amino acid change, and literature references are also included. | https://candl.osu.edu/ |
| DoCM (10.1038/nmeth.4000) | Curated repository from cancer variant databases and from individually curated publications | Curated mutations in cancer. | A database that aggregates, stores and tracks biologically important cancer variants. Literature references are included. | http://docm.info |
| Cancer Genome Interpreter (10.1186/s13073-018-0531-8) | Developed resources and computational methods | Annotations of tumor alterations and cancer genes. | Interprets, annotates, and analyzes cancer genomes across tumor types. Oncogenic alterations are identified and their potential impact on treatment response is also included. | https://www.cancergenomeinterpreter.org/ |
| OncoKB (10.1200/PO.17.00011) | Expert-guided knowledge base | Annotations of somatic molecular alterations in cancer-associated genes. | Contains annotations of more than 3,000 unique mutations, fusions, and copy number alterations in 418 cancer-associated genes. The biologic, oncogenic, prognostic and predictive effects and significance of the alterations are included. | http://oncokb.org |
| CancerTracer (10.1093/nar/gkz1061) | Manual curation | Evolutionary trajectories of tumor growth in individual patients. | They constructed patient-specific phylogenetic trees based on somatic mutations and copy number alterations. CancerTracer contains over 6000 tumor samples from 1548 patients across 45 different cancer types. | http://cailab.labshare.cn/cancertracer |
| FASMIC (10.1016/j.ccell.2018.01.021) | A combination of wet lab data in vivo and in silico methods | Functional impact of mutations. | Aims to assess the functional impact of somatic mutations and annotate these as "Activating", "Opposite", "Inactivating'", and "Neutral". | http://bioinformatics.mdanderson.org/main/FASMIC |

| Finding driver mutations in cancer: Elucidating the role of background mutational processes (10.1371/journal.pcbi.1006981) | Computational | Missense mutations from 58 genes. | This study created a dataset of missense mutations from 58 genes from different studies. This dataset comprises 5276 mutations (divided into 4137 neutral and 1139 non-neutral mutations) spanning the 58 genes. | https://www.ncbi.nlm.nih.gov/research/mutagene/benchmark |